\documentclass[%
 reprint,
 amsmath,amssymb,
]{revtex4-2}

\usepackage{physics}
\usepackage{hyperref}
\usepackage{graphicx}
\usepackage{dcolumn}
\usepackage{bm}
\usepackage{capt-of}

\DeclareMathOperator*{\argmin}{arg\,min}

\begin{document}

\preprint{APS/123-QED}

\title{Model-Agnostic Tagging of Quenched Jets in Heavy-Ion Collisions}

\author{Umar Sohail Qureshi}
 \email{uqureshi@cern.ch}%
\author{Raghav Kunnawalkam Elayavalli}

\affiliation{Department of Physics and Astronomy, Vanderbilt University, Nashville, Tennessee, USA}%

\date{\today}

\begin{abstract}
Measurements of jet substructure in ultra-relativistic heavy-ion collisions indicate that interactions with the quark-gluon plasma quench the jet showering process. Modern data-driven methods have shown promise in probing these modifications in the jet’s hard substructure. In this Letter, we present a machine learning framework to identify quenched jets while accounting for pileup, uncorrelated soft particle background, and detector effects---a more experimentally realistic and challenging scenario than previously addressed. Our approach leverages an interpretable sequential attention-based mechanism that integrates representations of individual jet constituents alongside global jet observables as features. The framework sets a new benchmark for tagging quenched jets with reduced model dependence.
\end{abstract}

\maketitle











\setlength{\parindent}{0pt}

\section{Introduction}
\label{introduction}
Ultra-relativistic heavy-ion collisions at CERN's Large Hadron Collider (LHC) and Brookhaven National Lab's Relativistic Heavy-Ion Collider (RHIC) offer a window into the quark-gluon plasma (QGP), a state of matter where asymptotically free quarks and gluons exist deconfined from hadronic matter. Among the most effective tools for probing the properties of the QGP are jets---collimated sprays of particles that emerge from the fragmentation of quarks and gluons produced in hard QCD scatterings \cite{Mehtar-Tani:2013pia}.

The phenomenon of jet quenching, where highly energetic partons are expected lose energy while traversing the QGP, has been extensively studied to understand the interaction between jets and the medium \cite{Bjorken:1982tu, dEnterria:2009xfs, Majumder:2010qh, Qin:2015srf}. For instance, studies of jet suppression \cite{PHENIX:2001hpc, STAR:2002ggv, ALICE:2010yje, PHENIX:2003qdj, STAR:2003fka, STAR:2002svs, ALICE:2011gpa, CMS:2012aa, ATLAS:2015qmb, CMS:2016xef}, dijet asymmetry \cite{ATLAS:2010isq, CMS:2011iwn, CMS:2012ulu, CMS:2015hkr}, and splitting functions \cite{CMS:2017qlm, STAR:2021kjt, ALargeIonColliderExperiment:2021mqf}, in addition to modifications in jet shapes \cite{CMS:2013lhm, ALICE:2019whv, ATLAS:2019pid}, fragmentation functions \cite{CMS:2014jjt, ATLAS:2017nre}, and energy-energy correlators \cite{Chen:2024cgx}, have provided insights into the redistribution of energy and momentum due to QGP interactions. Despite these advances, challenges remain in isolating quenched jets from unmodified ones due to the steeply falling jet spectrum and the presence of significant background contributions in heavy-ion environments.

The advent of machine learning (ML) techniques has opened new avenues for analyzing jet quenching \cite{hepmllivingreview}. Recent efforts have leveraged deep neural networks (NN) to classify quenched and unquenched jets \cite{Liu:2022hzd, Apolinario:2021olp, Lai:2021ckt, CrispimRomao:2023ssj, Du:2023qst}. However, existing methods often rely on idealized scenarios with minimal-to-no consideration of experimental effects such as pileup, uncorrelated soft particle background, and detector responses, limiting their applicability to real-world data.

In this Letter, we address these limitations by presenting a novel, highly interpretable machine learning framework that accurately identifies quenched jets under experimentally realistic conditions. Our approach employs an attention-based mechanism that integrates both global jet observables and the detailed substructure of individual jet constituents, enabling a more comprehensive understanding of jet-medium modifications. To ensure robustness and reduce model dependence, we utilize simulations from two different jet quenching models, \textsc{Jewel} \cite{Zapp:2012ak} and \textsc{CoLBT-Hydro} \cite{Chen:2017zte}, alongside non-quenching models, \textsc{Pythia}8 \cite{Sjostrand:2014zea, Sjostrand:2007gs} and \textsc{Herwig}7 \cite{Bewick:2023tfi, Bahr:2008pv}. This multi-model approach allows our framework to generalize across different theoretical descriptions of jet quenching.

Our method achieves a classification area under the curve (AUC) of 0.95, significantly outperforming state-of-the-art techniques, including CNNs, DeepSets, and LSTM-based models, which typically achieve AUC values between 0.75 and 0.85. This performance highlights the robustness of our framework in distinguishing quenched jets, even in the presence of complex experimental effects. Furthermore, by explicitly accounting for pileup and detector effects, our study bridges the gap between theoretical advancements and their experimental realization, setting a new benchmark for machine learning applications in heavy-ion physics.

The remainder of this Letter is organized as follows: Sec. \ref{sec:sampsim} describes the simulation setup and the procedure for generating and representing jets. In Sec. \ref{sec:ML}, we introduce the attention-based model architecture. Sec. \ref{sec:res} presents the results of our analysis, including comparisons with existing methods. We conclude with a discussion on the outlook of our findings for future experimental and theoretical studies in Sec. \ref{sec:disc}.


\section{Methodology}
\subsection{Samples and Simulation}
\label{sec:sampsim}

Samples are produced considering hard 2-2 QCD scattering events at $\sqrt{s_{\mathrm{NN}}}=5.02$ TeV with $\widehat{p}_\perp > 100$ GeV. Vacuum (pp) events are simulated using \textsc{Pythia8} using the Monash tune \cite{Skands:2014pea}, and \textsc{Herwig7} using the Nashville tune \cite{qureshi2024newherwig7underlyingevent}. Although, for the purposes of this study, no significant differences were observed between the results from the Nashville and the default tune. Events with the QGP medium (PbPb collisions) are generated using \textsc{Jewel} (with recoil) and \textsc{CoLBT-Hydro} with an initial temperature $T_i = 590$ MeV and initial quenching time $\tau_i = 0.4$, which provides an adequate description of a variety of jet quenching observables.

Following the guidelines in Ref. \cite{Andrews:2018jcm}, particles in the uncorrelated thermal background are simulated using a Boltzmann distribution for transverse momentum $p_\perp$ and a uniform distribution for pseudorapidity $\eta$ and azimuth $\phi$. We generate thermal background events corresponding to most-central (0–10\%) PbPb collisions at $\sqrt{s_\mathrm{NN}} = 5.02$ TeV, based on experimental data published in Ref. \cite{ALICE:2016fbt}. Details regarding the uncorrelated thermal background are summarized in Table \ref{tab:thermalbgsummary}. Pileup events are included utilizing minimum-bias simulations from \textsc{Pythia}8, with an average of 150 pileup interactions per event. These events represent additional minimum-bias collisions that occur in the same detector readout window as the primary hard scattering. 

Mixed events are subsequently constructed by the embedding dijet events into the underlying events, combining the hard scattering process with the soft thermal background and large pileup characteristic of heavy-ion collisions. Detector effects are simulated by smearing jet constituents. The $p_\perp$ smearing is modeled using a Gaussian distribution $\sim \mathcal{N}\left(p_\perp, (0.1p_\perp)^2\right)$ with a standard deviation proportional to the particle's $p_\perp$, reflecting the detector's energy-dependent resolution. Particles with $p_\perp < 0.5$ GeV are removed. Smearing in $\eta$ and $\phi$ is applied with fixed resolution $\sim \mathcal{N}\left(\eta/\phi, 0.1^2\right)$, simulating the angular resolution of the detector. Each jet constituent is treated independently, ensuring that the cumulative impact of smearing reflects realistic detector uncertainties.

To remove the contributions of thermal background and pileup particles from the jets, we apply background subtraction to the mixed events. In particular, we use the event-wide constituent subtraction method \cite{Berta:2014eza} with a maximum distance parameter $\Delta R_{\text{max}} = 0.3$ between signal and background particles, and a transverse momentum weight $\alpha = 1$. Particles that remain after subtraction are clustered into jets with the anti-$k_\perp$ algorithm \cite{Cacciari:2008gp} with a distance parameter $R = 0.4$ requiring $p_\perp^{\mathrm{jet}} > 100$ GeV and $\abs{\eta} < 2.5$ using \textsc{FastJet} \cite{Cacciari:2011ma, Cacciari:2005hq}. After completing the background subtraction and jet clustering, the final dataset includes approximately $\mathcal{O}(100\mathrm{k})$ jets per sample.

\begin{table}
    \centering
    \begin{tabular}{c|c}
        \hline
        \textbf{Quantity} & \textbf{Values} \\ \hline
        Pseudorapidity Interval & $|\eta| < 3$ \\
        Total Event Multiplicity & 7000 \\
        Corresponding Centrality Interval & 0--10\% \\
        Average Yield $\left\langle \dv{N}{\eta} \right\rangle$ & 1167 \\
        Average Transverse Momentum $\langle p_\perp \rangle$ & 1.2 GeV \\
        Average $p_\perp$ $\langle \rho \rangle$ & 227.1 GeV/Area \\
        \hline
    \end{tabular}
    \caption{Parameters of the simulated thermal background events.}
    \label{tab:thermalbgsummary}
\end{table}

Our workflow utilizes a vector representation of jets $\mathcal{J}$, integrating jet observables alongside jet constituents:
\begin{equation}
    \begin{aligned}
         \mathcal{J} = &\left\{n, m, p_\perp, z_g, R_g, k_\perp, m_g\right\}
\bigoplus_{i=1}^n \left\{{p}_\perp^{i}, \eta^i, \phi^i, \text{PID}^i\right\}
    \end{aligned}
\end{equation}
Where the $n$ is the particle multiplicity; $m$ is the (inclusive) invariant jet mass; $p_\perp$ is the jet transverse momentum; $(z_g, R_g)$ are the soft drop \cite{Larkoski:2014wba} groomed splitting function and jet radius respectively; $k_\perp$ is the transverse momentum scale; and $m_g$ is the soft drop groomed jet mass. We note that each of the groomed jet observables is taken from the final pair of hard branches that pass the soft drop condition. Before inputting the particles into the NN, we order them based on $p_\perp$ and ensure a fixed-length input by considering only the first 35 particles in the jet. If the jet contains fewer than 35 particles, the vector is padded with zeros. The particle IDs are smeared to reflect experimentally realistic particle ID information. Particles are classified as one of $\gamma, h^+, h^-, h^0, e^-, e^+, \mu^-, \mu^+$---represented as a value starting at 0 and incremented by 0.1 for each type. Finally, jets from the \textsc{Jewel} and \textsc{CoLBT-Hydro} samples are assigned a label of 1, while jets from the \textsc{Pythia} and \textsc{Herwig} samples are assigned a label of 0.

\subsection{Machine Learning Workflow}
\label{sec:ML}

A sequential attention-based NN \cite{arik2020tabnetattentiveinterpretabletabular} is used for this study. The effectiveness of BDTs and MLPs has been demonstrated in various experimental and phenomenological studies \cite{Ai:2022qvs, ATLAS:2017fak,  Arganda2024, flórez2024probinglightscalarsvectorlike}; MLPs consistently outperform BDT variants on tasks such as distinguishing Higgs boson processes from background in the canonical Higgs Boson dataset \cite{misc_higgs_280}, even when large ensembles are used. Sequential attention-based networks, however, surpass MLPs by achieving better performance with more compact representations \cite{arik2020tabnetattentiveinterpretabletabular, qureshi2024probingcompressedmassspectrum, ABCNet}, making them well-suited for our study. For a detailed discussion, we direct the interested reader to Ref. \cite{arik2020tabnetattentiveinterpretabletabular} but provide a brief overview of the salient points relevant to this work. The model employs sparse, instance-specific feature selection derived directly from the data and builds a sequential multi-step framework where each step adds to the decision-making process using the chosen features. The model takes an input ${f}\in\mathbb{R}^{B \times D}$ and performs instance-wise feature selection using a learnable mask $ {M}[i] \in \mathbb{R}^{B \times D} $, where $ B $ is the batch size, $ D $ is the feature dimension, and $ i $ is the decision step. The mask is computed as:
\begin{equation}
    {M}[i] = \mathrm{sparsemax}(P[i-1] \cdot h_i(a[i-1]))
\end{equation}
Where $h_i$ is a trainable function (in this case, a fully connected (FC) layer followed by batch normalization) applied to the output of the previous step $a[i-1] $ and $ P[i-1]$ is the prior scale term with $P[0]=1$, controlling feature reuse across steps and is given as:
\begin{equation}
     P[i] = \prod_{j=1}^{i} (\gamma - M[j])
\end{equation} 
Where $ \gamma$ is a relaxation parameter that determines how often features are reused across steps. The sparsemax activation ensures sparsity in the feature selection:
\begin{equation}
    \text{sparsemax}(z) = \argmin_{p \in \Delta^{D-1}} \| p - z \|^2
\end{equation}
Where $\Delta^{D-1}$ is the probability simplex. Once features are selected by $M[i]$, they are processed using a transformer to produce the representations  $d[i]$ and $a[i]$:
\begin{equation}
    [d[i], a[i]] = f_i(M[i] \cdot f)
\end{equation}
Where $ f \in \mathbb{R}^{B \times D} $ is the input feature matrix, $ d[i] \in \mathbb{R}^{B \times N_d} $ is the decision output for the current step $ i $, and $ a[i] \in \mathbb{R}^{B \times N_a} $ is the information passed to the next step. The final decision $d_\mathrm{out}$ is constructed as:
\begin{equation}
    d_{\mathrm{out}} = \sum_{i=1}^{N_{\mathrm{steps}}} \mathrm{ReLU}(d[i])
\end{equation}
Finally, a linear layer $W_{\mathrm{final}}d_{\mathrm{out}}$ gives us the model's output. The model's feature selection masks can provide insights into the importance of features chosen at each step. If $M_{bj} [i] = 0$, then $j^\mathrm{th}$ feature of the $b^\mathrm{th}$ has no contribution to the decision. The aggregate contribution of each decision step to the final decision is given as:
\begin{equation}
    \eta_b[i] = \sum_{c=1}^{N_d} \mathrm{ReLU}(d_{b,c}[i])
\end{equation}
Where $ \eta_b[i] $ denotes the aggregate contribution at the $ i^{\text{th}} $ decision step for the $ b^{\text{th}} $ sample. Intuitively, when $ d_{b,c}[i] < 0 $, the features at the $ i^{\text{th}} $ step do not contribute to the overall decision. Conversely, as $ \eta_b[i] $ increases, its role in the final decision becomes more significant. By scaling the decision mask at each step using $ \eta_b[i] $, we can define the aggregate feature importance mask as:
\begin{equation}
    M_{b,j}^{\text{agg}} = \frac{\displaystyle\sum_{i=1}^{N_{\text{steps}}} \eta_b[i] M_{b,j}[i]}{\displaystyle\sum_{j=1}^D \sum_{i=1}^{N_{\text{steps}}} \eta_b[i] M_{b,j}[i]}
    \label{eq:aggmask}
\end{equation}
Thereby allowing us to discriminate between the relative importance of features. 

\begin{figure}
    \centering
    \includegraphics[width=\linewidth]{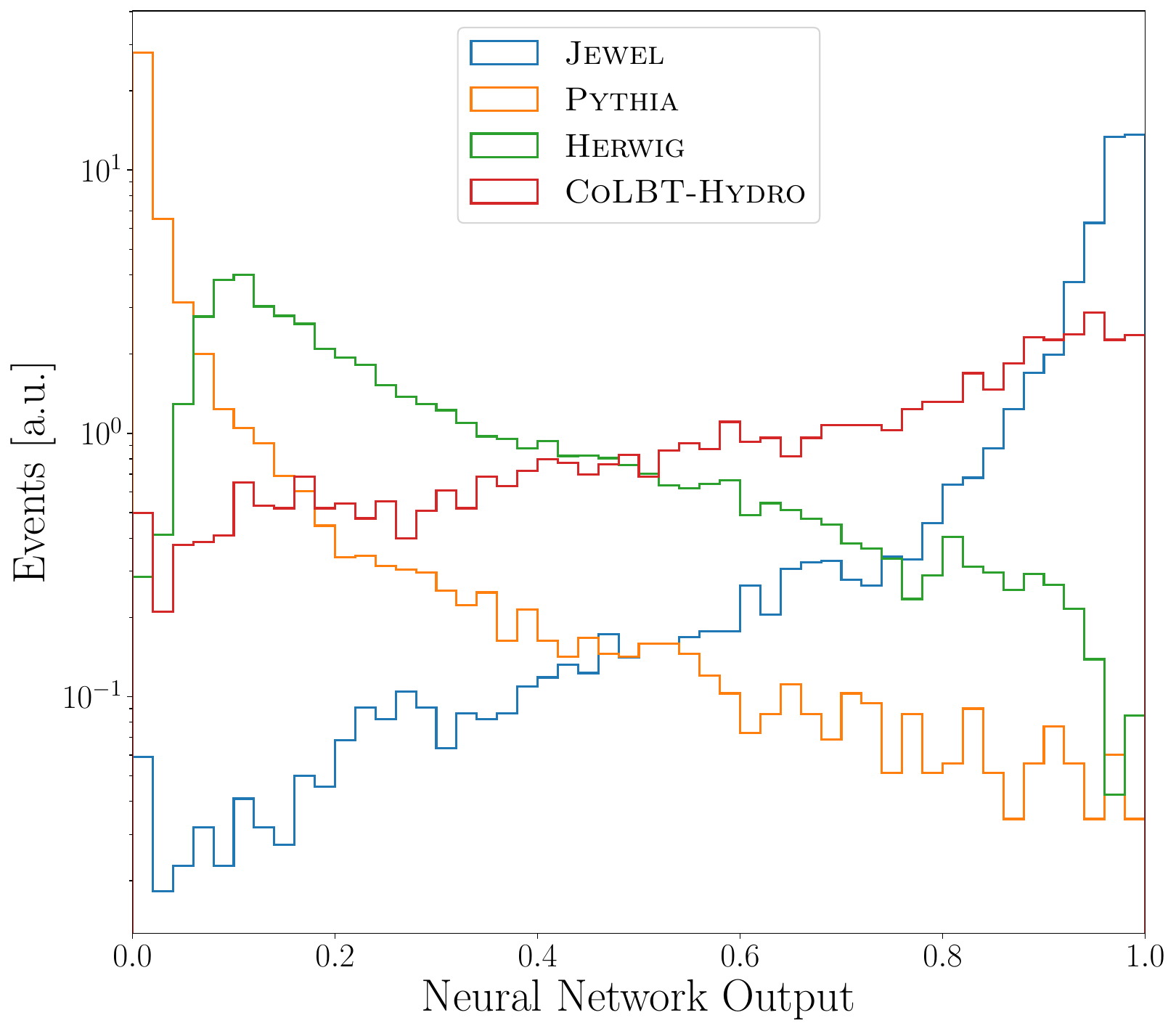}
    \caption{Distributions of the output for \textsc{Jewel} (blue), \textsc{Pythia} (orange), and \textsc{Herwig} (green), and \textsc{CoLBT-Hydro} (red) jets from the neural network. The distributions are normalized to unity.}
    \label{fig:Mloutdist}
\end{figure}

Training and evaluation of the model was performed out in a high-performance computing environment on an Nvidia T4 GPU. The canonical \textsc{PyTorch} \cite{paszke2019pytorchimperativestylehighperformance} deep learning framework was employed for implementing, configuring, training, and evaluating the models with a \textsc{Scikit-Learn} \cite{scikit-learn} wrapper for easy deployment. The model hyperparameters are tuned manually and are described in the remainder of this section. The size of the decision $n_d$ and attention layers $n_a$ is kept equal at $n_d=n_a = 8$; the number of steps $N_\mathrm{steps}=3$; the coefficient $\gamma$ for feature reusage in the masks is $\gamma = 1.3$. The Adam \cite{kingma2017adammethodstochasticoptimization} optimizer with a learning rate $\alpha = 0.02$ is used in conjunction with the mean-squared error loss function. A train-test split of 90-10 is used with a batch size of $16384$ for training. Early stopping is used with a patience of 10 epochs. 

\section{Results}
\begin{table*}
\centering
\begin{tabular}{c c c c c c c}
\hline
\textbf{Model}~~ & \textbf{Thermal Background}~~ & \textbf{Detector Effects}~~ & \textbf{Pileup}~~ & \textbf{Performance (AUC)} ~~& \textbf{Reference}  \\ \hline
Energy Flow Network & $\cross$ & $\cross$ & $\cross$  & 0.67 & \cite{Lai:2021ckt} \\ 
Particle Flow Network & $\cross$ & $\cross$ & $\cross$ & 0.86 & \cite{Lai:2021ckt} \\ 
Particle Flow Network & $\checkmark$ & $\cross$ & $\cross$  & 
 0.75 & \cite{Lai:2021ckt}\\ 
Long-Short Term Memory & \checkmark & $\cross$ & $\cross$  & 0.76  & \cite{Liu:2022hzd} \\ 
Long-Short Term Memory & \checkmark & $\cross$ & $\cross$  & 0.74  & \cite{Apolinario:2021olp} \\ 
Multi-Layer Perceptron & \checkmark & $\cross$ & $\cross$  & 0.73  & \cite{Apolinario:2021olp} \\ 
Autoencoder + Decision Tree & \checkmark & $\cross$ & $\cross$  & 0.70  & \cite{CrispimRomao:2023ssj} \\ 
Convolutional NN & \checkmark & $\cross$ & $\cross$  & 0.75  & \cite{Apolinario:2021olp}  \\ \hline
\textbf{Sequential Attention} & \checkmark & \checkmark & \checkmark  & \textbf{0.95} & Our Study  \\ 
\hline 
\end{tabular}
\caption{Performance comparison of existing machine learning approaches for quenched jet tagging under varying experimental conditions. The sequential attention-based model significantly outperforms all other methods while taking on a more challenging classification task.}
\label{tab:compres}
\end{table*}
\label{sec:res}
\begin{figure}
    \centering
    \includegraphics[width=0.95\linewidth]{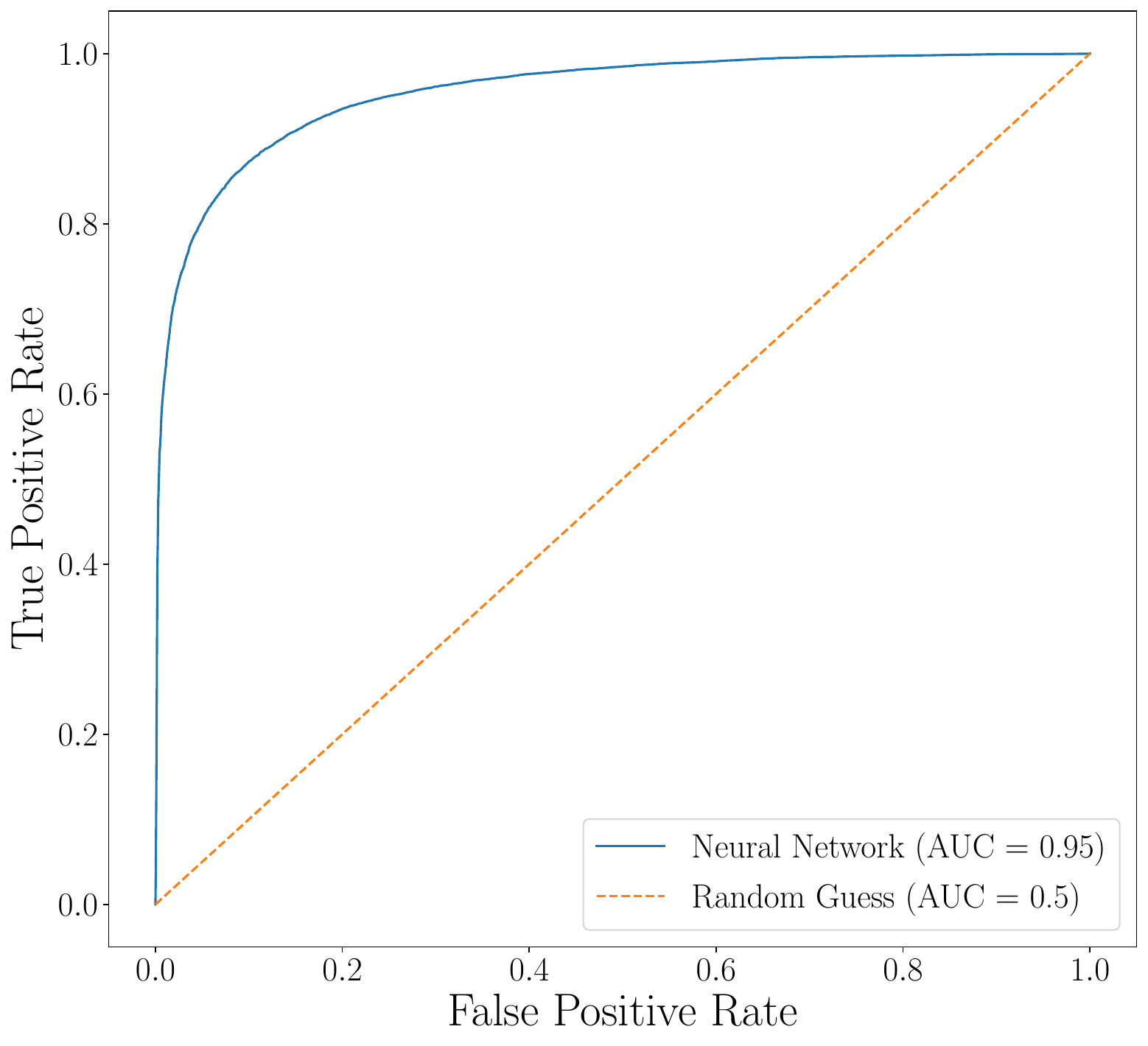}
    \caption{ROC curve for the separation of the vacuum and medium jets.}
    \label{fig:ROC}
\end{figure}
Figure \ref{fig:Mloutdist} shows the model output distributions for jets simulated using \textsc{Jewel}, \textsc{CoLBT-Hydro}, \textsc{Pythia}, and \textsc{Herwig}. The distributions reveal distinct, order-of-magnitude separation between quenched and vacuum jets, highlighting the model's ability to differentiate between them.

Figure \ref{fig:ROC} presents the receiver operating characteristic (ROC) curve for the classification task, achieving an area under the curve (AUC) of 0.95. Table \ref{tab:compres} compares the AUC values of our method with those of prior approaches. While existing methods achieve AUCs in the range of 0.67 to 0.86, our sequential attention framework excels while also accurately incorporating and addressing thermal background, detector effects, and pileup, setting a new benchmark for machine learning applications in quenched jet tagging. 

\begin{figure*}
    \centering
    \begin{minipage}{0.495\linewidth}
        \includegraphics[width=\linewidth]{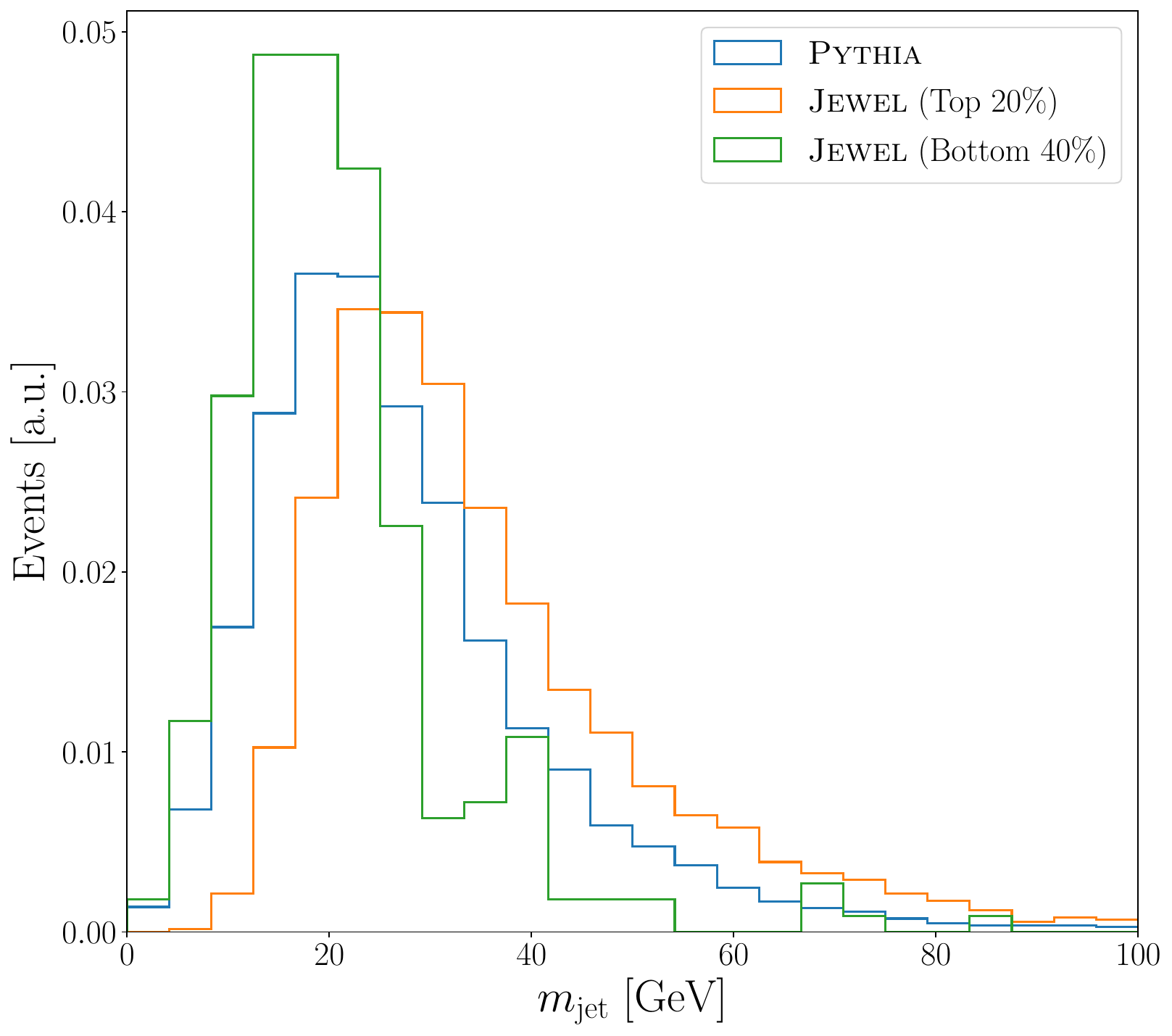}
    \end{minipage}
    \begin{minipage}{0.495\linewidth}
        \includegraphics[width=\linewidth]{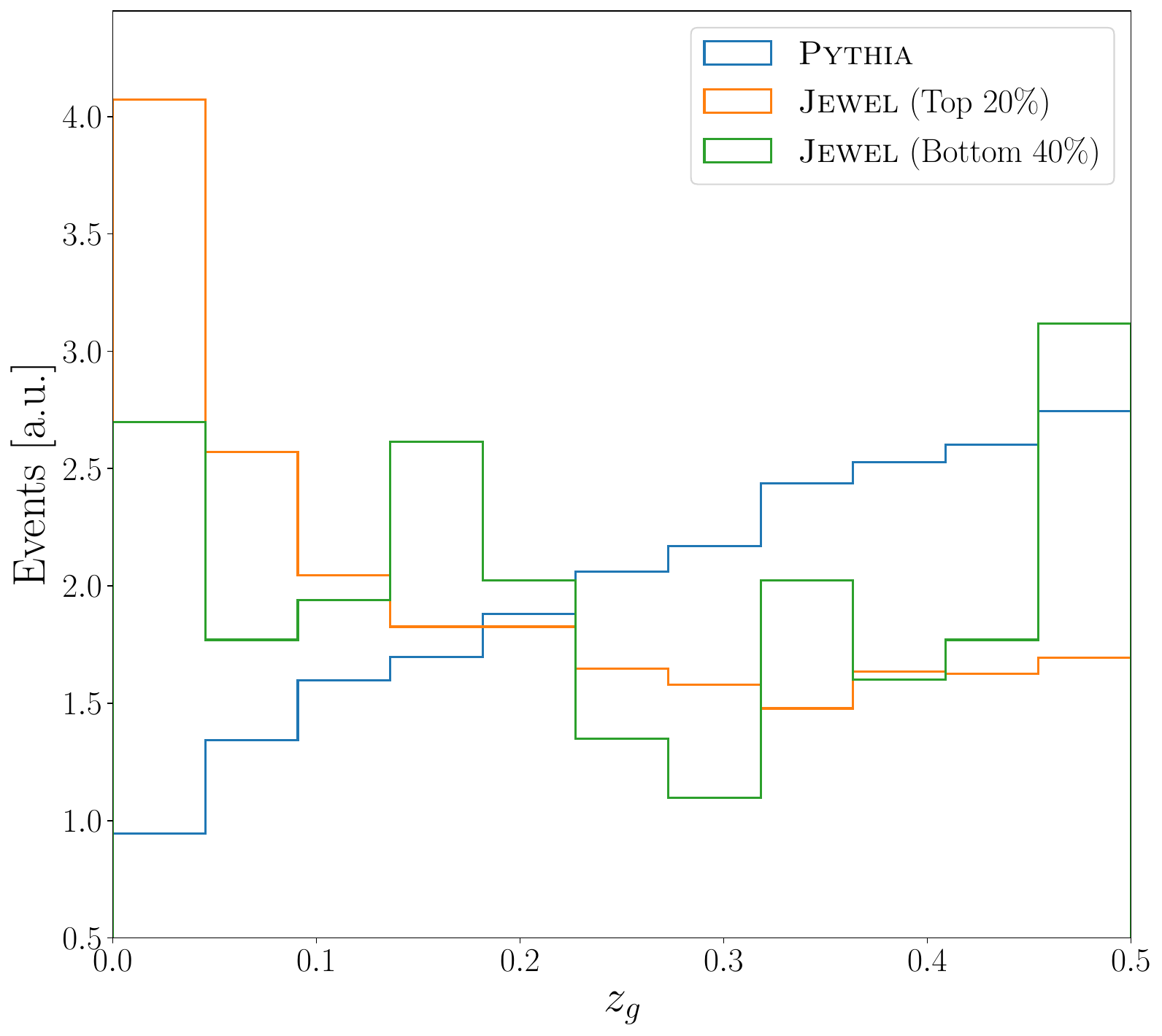}
    \end{minipage}
    \caption{Distributions of the inclusive mass $m_\mathrm{jet}$ (left) and the soft drop groomed splitting function $z_g$ (right) for the top 20\% (orange) and bottom 40\% (green) of \textsc{Jewel} jets. Sample \textsc{Pythia} (blue) jets are also shown as a baseline.}
    \label{fig:jetdists}
\end{figure*}

To investigate the relationship between the NN output and quenching effects, we split the \textsc{Jewel} samples into two subsets: top 20\% (NN output $> 0.8$) and bottom 40\% (NN output $< 0.4$). The jet mass and splitting function distributions in Fig. \ref{fig:jetdists} show significant modifications for the top 20\% while the bottom 40\% exhibits a quenching pattern closely resembling that of the \textsc{Pythia} jets.

\begin{figure*}
    \centering
    \begin{minipage}{0.495\linewidth}
        \includegraphics[width=\linewidth]{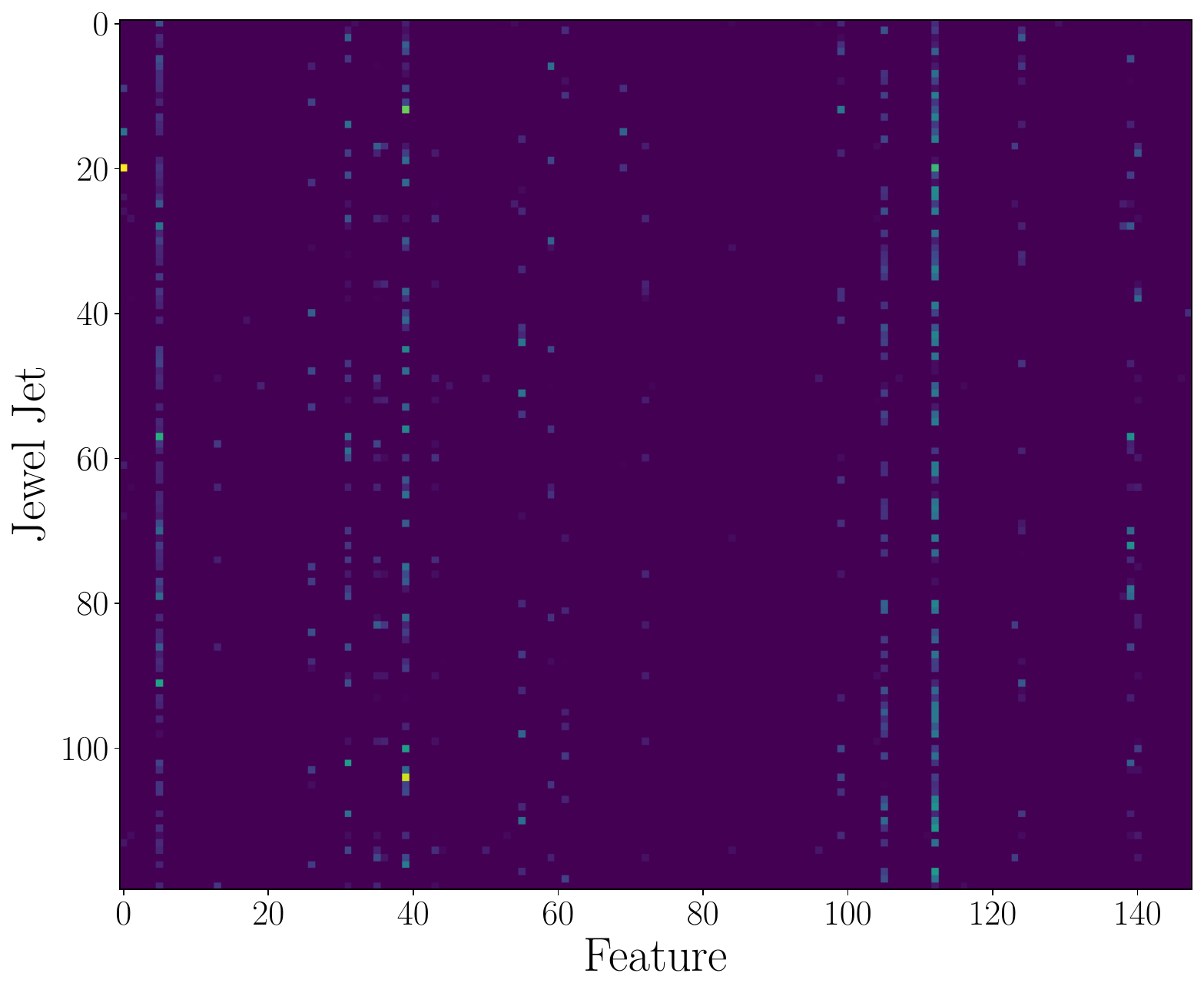}
    \end{minipage}
    \begin{minipage}{0.495\linewidth}
        \includegraphics[width=\linewidth]{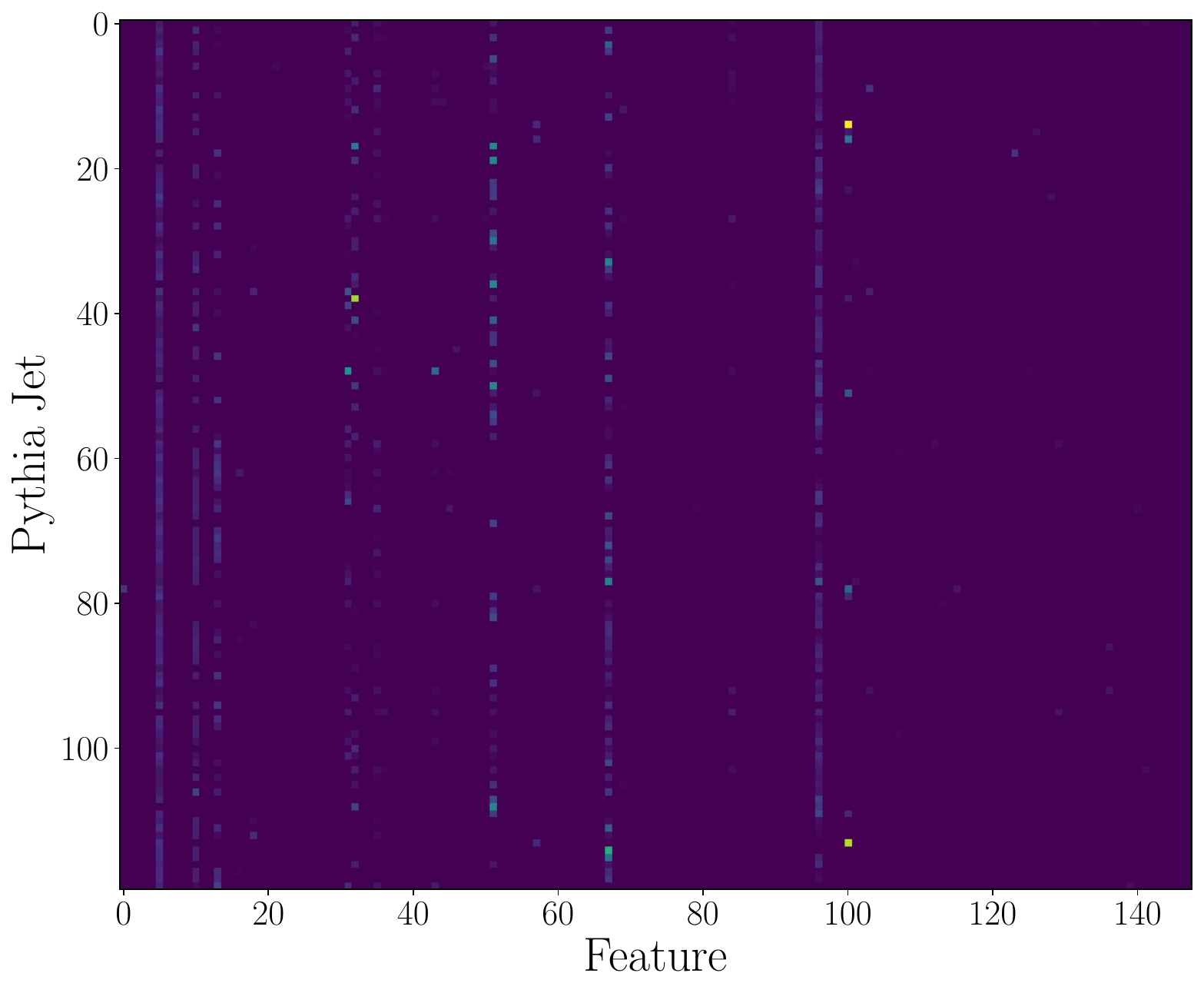}
    \end{minipage}
    \caption{Heatmap illustrations of the aggregate feature mask (Eq. \ref{eq:aggmask}) for the first 125 \textsc{Jewel} (left) and \textsc{Pythia} (right) truth jets. The sparsity in feature activation highlights the attention-based mechanism's focus on relevant features for classification.}
    \label{fig:masks}
\end{figure*}

In addition to superior performance, the interpretability of our model offers a unique advantage. By leveraging sparse attention-based feature selection, the model identifies the most critical features contributing to its decisions at each step. Analysis of the aggregate feature mask for different input samples, illustrated in Fig. \ref{fig:masks}, reveals consistent trends in sparsity, with only a subset of features being activated. These masks shed light on the specific characteristics most relevant to the classification task. Notably, features corresponding to the highest and lowest $p_\perp$ jet constituents play the biggest roles, indicative of energy redistribution and medium-induced modifications.

This sparsity not only enhances the computational efficiency of the model but also paves the way for ML-assisted observable design. By identifying the features most influential in tagging quenched jets, the feature masks can guide the development of new observables tailored to probe jet-medium interactions. While designing such an observable is well beyond the scope of this study, it is conceivable that our method can bridge data-driven approaches with traditional physics-driven methodologies. 


\section{Discussion}
\label{sec:disc}
This study introduces the first model-agnostic, interpretable machine learning framework to distinguish between quenched and unquenched jets under realistic experimental conditions, including thermal background, detector effects, and pileup. The sequential attention mechanism employed by the model significantly enhances its performance, achieving an AUC of 0.95 and outperforming existing methods. The framework utilizes a jet representation that integrates both global jet observables and the jet substructure.

Another key contribution of this study is the model's interpretability through its feature selection masks, which can provide insights into the physics of jet quenching. The sparsity in these masks highlights the efficiency of the attention mechanism in focusing on the most relevant features while irrelevant information. This not only aids in improving the classification performance but also opens new avenues for designing novel, physics-informed jet substructure observables. 

Our results bridge the gap between generator-level modeling and experimental measurements by incorporating experimental realities like pileup and detector effects, which are often overlooked in existing studies. As such, this work sets a new benchmark for machine learning applications in heavy-ion physics, demonstrating the potential of interpretable ML models to provide both high accuracy and valuable physical insights.

\pagebreak
$$ $$
\section*{Data and Code Availability}
All code and data necessary to reproduce the results presented in this study are publicly available on GitHub at \url{https://github.com/umarsqureshi/jetQuenchingML}. Detailed documentation is provided to facilitate reproducibility and further exploration by the community.

\begin{acknowledgments}
The authors express their gratitude to Y. Wu, J. Velkovska, and Y. (Luna) Chen for insightful discussions. Special thanks are extended to Y. Wu for supplying datasets utilized in the earlier phases of this study and Z. Yang for providing the \textsc{CoLBT-Hydro} samples. The authors also appreciate the support of the Vanderbilt ACCRE computing facility. RKE acknowledges funding from the U.S. Department of Energy, Office of Science, Office of Nuclear Physics under grant DE-SC0024660. USQ acknowledges funding from Immersion Vanderbilt.

\end{acknowledgments}



\bibliography{apssamp}

\begin{thebibliography}{60}%
\makeatletter
\providecommand \@ifxundefined [1]{%
 \@ifx{#1\undefined}
}%
\providecommand \@ifnum [1]{%
 \ifnum #1\expandafter \@firstoftwo
 \else \expandafter \@secondoftwo
 \fi
}%
\providecommand \@ifx [1]{%
 \ifx #1\expandafter \@firstoftwo
 \else \expandafter \@secondoftwo
 \fi
}%
\providecommand \natexlab [1]{#1}%
\providecommand \enquote  [1]{``#1''}%
\providecommand \bibnamefont  [1]{#1}%
\providecommand \bibfnamefont [1]{#1}%
\providecommand \citenamefont [1]{#1}%
\providecommand \href@noop [0]{\@secondoftwo}%
\providecommand \href [0]{\begingroup \@sanitize@url \@href}%
\providecommand \@href[1]{\@@startlink{#1}\@@href}%
\providecommand \@@href[1]{\endgroup#1\@@endlink}%
\providecommand \@sanitize@url [0]{\catcode `\\12\catcode `\$12\catcode `\&12\catcode `\#12\catcode `\^12\catcode `\_12\catcode `\%12\relax}%
\providecommand \@@startlink[1]{}%
\providecommand \@@endlink[0]{}%
\providecommand \url  [0]{\begingroup\@sanitize@url \@url }%
\providecommand \@url [1]{\endgroup\@href {#1}{\urlprefix }}%
\providecommand \urlprefix  [0]{URL }%
\providecommand \Eprint [0]{\href }%
\providecommand \doibase [0]{https://doi.org/}%
\providecommand \selectlanguage [0]{\@gobble}%
\providecommand \bibinfo  [0]{\@secondoftwo}%
\providecommand \bibfield  [0]{\@secondoftwo}%
\providecommand \translation [1]{[#1]}%
\providecommand \BibitemOpen [0]{}%
\providecommand \bibitemStop [0]{}%
\providecommand \bibitemNoStop [0]{.\EOS\space}%
\providecommand \EOS [0]{\spacefactor3000\relax}%
\providecommand \BibitemShut  [1]{\csname bibitem#1\endcsname}%
\let\auto@bib@innerbib\@empty
\bibitem [{\citenamefont {Mehtar-Tani}\ \emph {et~al.}(2013)\citenamefont {Mehtar-Tani}, \citenamefont {Milhano},\ and\ \citenamefont {Tywoniuk}}]{Mehtar-Tani:2013pia}%
  \BibitemOpen
  \bibfield  {author} {\bibinfo {author} {\bibfnamefont {Y.}~\bibnamefont {Mehtar-Tani}}, \bibinfo {author} {\bibfnamefont {J.~G.}\ \bibnamefont {Milhano}},\ and\ \bibinfo {author} {\bibfnamefont {K.}~\bibnamefont {Tywoniuk}},\ }\bibfield  {title} {\bibinfo {title} {{Jet physics in heavy-ion collisions}},\ }\href {https://doi.org/10.1142/S0217751X13400137} {\bibfield  {journal} {\bibinfo  {journal} {Int. J. Mod. Phys. A}\ }\textbf {\bibinfo {volume} {28}},\ \bibinfo {pages} {1340013} (\bibinfo {year} {2013})},\ \Eprint {https://arxiv.org/abs/1302.2579} {arXiv:1302.2579 [hep-ph]} \BibitemShut {NoStop}%
\bibitem [{\citenamefont {Bjorken}(1982)}]{Bjorken:1982tu}%
  \BibitemOpen
  \bibfield  {author} {\bibinfo {author} {\bibfnamefont {J.~D.}\ \bibnamefont {Bjorken}},\ }\href {https://cds.cern.ch/record/141477} {\emph {\bibinfo {title} {{Energy loss of energetic partons in quark-gluon plasma: possible extinction of high p$_{T}$ jets in hadron-hadron collisions}}}},\ \bibinfo {type} {Tech. Rep.}\ (\bibinfo  {institution} {FERMILAB},\ \bibinfo {address} {Batavia, IL},\ \bibinfo {year} {1982})\BibitemShut {NoStop}%
\bibitem [{\citenamefont {d'Enterria}(2010)}]{dEnterria:2009xfs}%
  \BibitemOpen
  \bibfield  {author} {\bibinfo {author} {\bibfnamefont {D.}~\bibnamefont {d'Enterria}},\ }\bibfield  {title} {\bibinfo {title} {{Jet quenching}},\ }\href {https://doi.org/10.1007/978-3-642-01539-7_16} {\bibfield  {journal} {\bibinfo  {journal} {Landolt-Bornstein}\ }\textbf {\bibinfo {volume} {23}},\ \bibinfo {pages} {471} (\bibinfo {year} {2010})},\ \Eprint {https://arxiv.org/abs/0902.2011} {arXiv:0902.2011 [nucl-ex]} \BibitemShut {NoStop}%
\bibitem [{\citenamefont {Majumder}\ and\ \citenamefont {Van~Leeuwen}(2011)}]{Majumder:2010qh}%
  \BibitemOpen
  \bibfield  {author} {\bibinfo {author} {\bibfnamefont {A.}~\bibnamefont {Majumder}}\ and\ \bibinfo {author} {\bibfnamefont {M.}~\bibnamefont {Van~Leeuwen}},\ }\bibfield  {title} {\bibinfo {title} {{The Theory and Phenomenology of Perturbative QCD Based Jet Quenching}},\ }\href {https://doi.org/10.1016/j.ppnp.2010.09.001} {\bibfield  {journal} {\bibinfo  {journal} {Prog. Part. Nucl. Phys.}\ }\textbf {\bibinfo {volume} {66}},\ \bibinfo {pages} {41} (\bibinfo {year} {2011})},\ \Eprint {https://arxiv.org/abs/1002.2206} {arXiv:1002.2206 [hep-ph]} \BibitemShut {NoStop}%
\bibitem [{\citenamefont {Qin}\ and\ \citenamefont {Wang}(2015)}]{Qin:2015srf}%
  \BibitemOpen
  \bibfield  {author} {\bibinfo {author} {\bibfnamefont {G.-Y.}\ \bibnamefont {Qin}}\ and\ \bibinfo {author} {\bibfnamefont {X.-N.}\ \bibnamefont {Wang}},\ }\bibfield  {title} {\bibinfo {title} {{Jet quenching in high-energy heavy-ion collisions}},\ }\href {https://doi.org/10.1142/S0218301315300143} {\bibfield  {journal} {\bibinfo  {journal} {Int. J. Mod. Phys. E}\ }\textbf {\bibinfo {volume} {24}},\ \bibinfo {pages} {1530014} (\bibinfo {year} {2015})},\ \Eprint {https://arxiv.org/abs/1511.00790} {arXiv:1511.00790 [hep-ph]} \BibitemShut {NoStop}%
\bibitem [{\citenamefont {Adcox}\ \emph {et~al.}(2002)\citenamefont {Adcox} \emph {et~al.}}]{PHENIX:2001hpc}%
  \BibitemOpen
  \bibfield  {author} {\bibinfo {author} {\bibfnamefont {K.}~\bibnamefont {Adcox}} \emph {et~al.} (\bibinfo {collaboration} {PHENIX}),\ }\bibfield  {title} {\bibinfo {title} {{Suppression of hadrons with large transverse momentum in central Au+Au collisions at $\sqrt{s_{NN}}$ = 130-GeV}},\ }\href {https://doi.org/10.1103/PhysRevLett.88.022301} {\bibfield  {journal} {\bibinfo  {journal} {Phys. Rev. Lett.}\ }\textbf {\bibinfo {volume} {88}},\ \bibinfo {pages} {022301} (\bibinfo {year} {2002})},\ \Eprint {https://arxiv.org/abs/nucl-ex/0109003} {arXiv:nucl-ex/0109003} \BibitemShut {NoStop}%
\bibitem [{\citenamefont {Adler}\ \emph {et~al.}(2002)\citenamefont {Adler} \emph {et~al.}}]{STAR:2002ggv}%
  \BibitemOpen
  \bibfield  {author} {\bibinfo {author} {\bibfnamefont {C.}~\bibnamefont {Adler}} \emph {et~al.} (\bibinfo {collaboration} {STAR}),\ }\bibfield  {title} {\bibinfo {title} {{Centrality dependence of high $p_{T}$ hadron suppression in Au+Au collisions at $\sqrt{s}_{NN}$ = 130-GeV}},\ }\href {https://doi.org/10.1103/PhysRevLett.89.202301} {\bibfield  {journal} {\bibinfo  {journal} {Phys. Rev. Lett.}\ }\textbf {\bibinfo {volume} {89}},\ \bibinfo {pages} {202301} (\bibinfo {year} {2002})},\ \Eprint {https://arxiv.org/abs/nucl-ex/0206011} {arXiv:nucl-ex/0206011} \BibitemShut {NoStop}%
\bibitem [{\citenamefont {Aamodt}\ \emph {et~al.}(2011)\citenamefont {Aamodt} \emph {et~al.}}]{ALICE:2010yje}%
  \BibitemOpen
  \bibfield  {author} {\bibinfo {author} {\bibfnamefont {K.}~\bibnamefont {Aamodt}} \emph {et~al.} (\bibinfo {collaboration} {ALICE}),\ }\bibfield  {title} {\bibinfo {title} {{Suppression of Charged Particle Production at Large Transverse Momentum in Central Pb-Pb Collisions at $\sqrt{s_{NN}} =$ 2.76 TeV}},\ }\href {https://doi.org/10.1016/j.physletb.2010.12.020} {\bibfield  {journal} {\bibinfo  {journal} {Phys. Lett. B}\ }\textbf {\bibinfo {volume} {696}},\ \bibinfo {pages} {30} (\bibinfo {year} {2011})},\ \Eprint {https://arxiv.org/abs/1012.1004} {arXiv:1012.1004 [nucl-ex]} \BibitemShut {NoStop}%
\bibitem [{\citenamefont {Adler}\ \emph {et~al.}(2003{\natexlab{a}})\citenamefont {Adler} \emph {et~al.}}]{PHENIX:2003qdj}%
  \BibitemOpen
  \bibfield  {author} {\bibinfo {author} {\bibfnamefont {S.~S.}\ \bibnamefont {Adler}} \emph {et~al.} (\bibinfo {collaboration} {PHENIX}),\ }\bibfield  {title} {\bibinfo {title} {{Suppressed $\pi^0$ production at large transverse momentum in central Au+ Au collisions at $\sqrt{S_{NN}}$ = 200 GeV}},\ }\href {https://doi.org/10.1103/PhysRevLett.91.072301} {\bibfield  {journal} {\bibinfo  {journal} {Phys. Rev. Lett.}\ }\textbf {\bibinfo {volume} {91}},\ \bibinfo {pages} {072301} (\bibinfo {year} {2003}{\natexlab{a}})},\ \Eprint {https://arxiv.org/abs/nucl-ex/0304022} {arXiv:nucl-ex/0304022} \BibitemShut {NoStop}%
\bibitem [{\citenamefont {Adams}\ \emph {et~al.}(2003)\citenamefont {Adams} \emph {et~al.}}]{STAR:2003fka}%
  \BibitemOpen
  \bibfield  {author} {\bibinfo {author} {\bibfnamefont {J.}~\bibnamefont {Adams}} \emph {et~al.} (\bibinfo {collaboration} {STAR}),\ }\bibfield  {title} {\bibinfo {title} {{Transverse momentum and collision energy dependence of high p(T) hadron suppression in Au+Au collisions at ultrarelativistic energies}},\ }\href {https://doi.org/10.1103/PhysRevLett.91.172302} {\bibfield  {journal} {\bibinfo  {journal} {Phys. Rev. Lett.}\ }\textbf {\bibinfo {volume} {91}},\ \bibinfo {pages} {172302} (\bibinfo {year} {2003})},\ \Eprint {https://arxiv.org/abs/nucl-ex/0305015} {arXiv:nucl-ex/0305015} \BibitemShut {NoStop}%
\bibitem [{\citenamefont {Adler}\ \emph {et~al.}(2003{\natexlab{b}})\citenamefont {Adler} \emph {et~al.}}]{STAR:2002svs}%
  \BibitemOpen
  \bibfield  {author} {\bibinfo {author} {\bibfnamefont {C.}~\bibnamefont {Adler}} \emph {et~al.} (\bibinfo {collaboration} {STAR}),\ }\bibfield  {title} {\bibinfo {title} {{Disappearance of back-to-back high $p_{T}$ hadron correlations in central Au+Au collisions at $\sqrt{s_{NN}}$ = 200-GeV}},\ }\href {https://doi.org/10.1103/PhysRevLett.90.082302} {\bibfield  {journal} {\bibinfo  {journal} {Phys. Rev. Lett.}\ }\textbf {\bibinfo {volume} {90}},\ \bibinfo {pages} {082302} (\bibinfo {year} {2003}{\natexlab{b}})},\ \Eprint {https://arxiv.org/abs/nucl-ex/0210033} {arXiv:nucl-ex/0210033} \BibitemShut {NoStop}%
\bibitem [{\citenamefont {Aamodt}\ \emph {et~al.}(2012)\citenamefont {Aamodt} \emph {et~al.}}]{ALICE:2011gpa}%
  \BibitemOpen
  \bibfield  {author} {\bibinfo {author} {\bibfnamefont {K.}~\bibnamefont {Aamodt}} \emph {et~al.} (\bibinfo {collaboration} {ALICE}),\ }\bibfield  {title} {\bibinfo {title} {{Particle-yield modification in jet-like azimuthal di-hadron correlations in Pb-Pb collisions at $\sqrt{s_{NN}} = 2.76$ TeV}},\ }\href {https://doi.org/10.1103/PhysRevLett.108.092301} {\bibfield  {journal} {\bibinfo  {journal} {Phys. Rev. Lett.}\ }\textbf {\bibinfo {volume} {108}},\ \bibinfo {pages} {092301} (\bibinfo {year} {2012})},\ \Eprint {https://arxiv.org/abs/1110.0121} {arXiv:1110.0121 [nucl-ex]} \BibitemShut {NoStop}%
\bibitem [{\citenamefont {Chatrchyan}\ \emph {et~al.}(2012{\natexlab{a}})\citenamefont {Chatrchyan} \emph {et~al.}}]{CMS:2012aa}%
  \BibitemOpen
  \bibfield  {author} {\bibinfo {author} {\bibfnamefont {S.}~\bibnamefont {Chatrchyan}} \emph {et~al.} (\bibinfo {collaboration} {CMS}),\ }\bibfield  {title} {\bibinfo {title} {{Study of High-pT Charged Particle Suppression in PbPb Compared to $pp$ Collisions at $\sqrt{s_{NN}}=2.76$ TeV}},\ }\href {https://doi.org/10.1140/epjc/s10052-012-1945-x} {\bibfield  {journal} {\bibinfo  {journal} {Eur. Phys. J. C}\ }\textbf {\bibinfo {volume} {72}},\ \bibinfo {pages} {1945} (\bibinfo {year} {2012}{\natexlab{a}})},\ \Eprint {https://arxiv.org/abs/1202.2554} {arXiv:1202.2554 [nucl-ex]} \BibitemShut {NoStop}%
\bibitem [{\citenamefont {Aad}\ \emph {et~al.}(2015)\citenamefont {Aad} \emph {et~al.}}]{ATLAS:2015qmb}%
  \BibitemOpen
  \bibfield  {author} {\bibinfo {author} {\bibfnamefont {G.}~\bibnamefont {Aad}} \emph {et~al.} (\bibinfo {collaboration} {ATLAS}),\ }\bibfield  {title} {\bibinfo {title} {{Measurement of charged-particle spectra in Pb+Pb collisions at $\sqrt{{s}_\mathsf{{NN}}} = 2.76$ TeV with the ATLAS detector at the LHC}},\ }\href {https://doi.org/10.1007/JHEP09(2015)050} {\bibfield  {journal} {\bibinfo  {journal} {JHEP}\ }\textbf {\bibinfo {volume} {09}},\ \bibinfo {pages} {050}},\ \Eprint {https://arxiv.org/abs/1504.04337} {arXiv:1504.04337 [hep-ex]} \BibitemShut {NoStop}%
\bibitem [{\citenamefont {Khachatryan}\ \emph {et~al.}(2017)\citenamefont {Khachatryan} \emph {et~al.}}]{CMS:2016xef}%
  \BibitemOpen
  \bibfield  {author} {\bibinfo {author} {\bibfnamefont {V.}~\bibnamefont {Khachatryan}} \emph {et~al.} (\bibinfo {collaboration} {CMS}),\ }\bibfield  {title} {\bibinfo {title} {{Charged-particle nuclear modification factors in PbPb and pPb collisions at $ \sqrt{s_{\mathrm{N}\;\mathrm{N}}}=5.02 $ TeV}},\ }\href {https://doi.org/10.1007/JHEP04(2017)039} {\bibfield  {journal} {\bibinfo  {journal} {JHEP}\ }\textbf {\bibinfo {volume} {04}},\ \bibinfo {pages} {039}},\ \Eprint {https://arxiv.org/abs/1611.01664} {arXiv:1611.01664 [nucl-ex]} \BibitemShut {NoStop}%
\bibitem [{\citenamefont {Aad}\ \emph {et~al.}(2010)\citenamefont {Aad} \emph {et~al.}}]{ATLAS:2010isq}%
  \BibitemOpen
  \bibfield  {author} {\bibinfo {author} {\bibfnamefont {G.}~\bibnamefont {Aad}} \emph {et~al.} (\bibinfo {collaboration} {ATLAS}),\ }\bibfield  {title} {\bibinfo {title} {{Observation of a Centrality-Dependent Dijet Asymmetry in Lead-Lead Collisions at $\sqrt{s_{NN}}=2.77$ TeV with the ATLAS Detector at the LHC}},\ }\href {https://doi.org/10.1103/PhysRevLett.105.252303} {\bibfield  {journal} {\bibinfo  {journal} {Phys. Rev. Lett.}\ }\textbf {\bibinfo {volume} {105}},\ \bibinfo {pages} {252303} (\bibinfo {year} {2010})},\ \Eprint {https://arxiv.org/abs/1011.6182} {arXiv:1011.6182 [hep-ex]} \BibitemShut {NoStop}%
\bibitem [{\citenamefont {Chatrchyan}\ \emph {et~al.}(2011)\citenamefont {Chatrchyan} \emph {et~al.}}]{CMS:2011iwn}%
  \BibitemOpen
  \bibfield  {author} {\bibinfo {author} {\bibfnamefont {S.}~\bibnamefont {Chatrchyan}} \emph {et~al.} (\bibinfo {collaboration} {CMS}),\ }\bibfield  {title} {\bibinfo {title} {{Observation and studies of jet quenching in PbPb collisions at nucleon-nucleon center-of-mass energy = 2.76 TeV}},\ }\href {https://doi.org/10.1103/PhysRevC.84.024906} {\bibfield  {journal} {\bibinfo  {journal} {Phys. Rev. C}\ }\textbf {\bibinfo {volume} {84}},\ \bibinfo {pages} {024906} (\bibinfo {year} {2011})},\ \Eprint {https://arxiv.org/abs/1102.1957} {arXiv:1102.1957 [nucl-ex]} \BibitemShut {NoStop}%
\bibitem [{\citenamefont {Chatrchyan}\ \emph {et~al.}(2012{\natexlab{b}})\citenamefont {Chatrchyan} \emph {et~al.}}]{CMS:2012ulu}%
  \BibitemOpen
  \bibfield  {author} {\bibinfo {author} {\bibfnamefont {S.}~\bibnamefont {Chatrchyan}} \emph {et~al.} (\bibinfo {collaboration} {CMS}),\ }\bibfield  {title} {\bibinfo {title} {{Jet Momentum Dependence of Jet Quenching in PbPb Collisions at $\sqrt{s_{NN}}=2.76$ TeV}},\ }\href {https://doi.org/10.1016/j.physletb.2012.04.058} {\bibfield  {journal} {\bibinfo  {journal} {Phys. Lett. B}\ }\textbf {\bibinfo {volume} {712}},\ \bibinfo {pages} {176} (\bibinfo {year} {2012}{\natexlab{b}})},\ \Eprint {https://arxiv.org/abs/1202.5022} {arXiv:1202.5022 [nucl-ex]} \BibitemShut {NoStop}%
\bibitem [{\citenamefont {Khachatryan}\ \emph {et~al.}(2016)\citenamefont {Khachatryan} \emph {et~al.}}]{CMS:2015hkr}%
  \BibitemOpen
  \bibfield  {author} {\bibinfo {author} {\bibfnamefont {V.}~\bibnamefont {Khachatryan}} \emph {et~al.} (\bibinfo {collaboration} {CMS}),\ }\bibfield  {title} {\bibinfo {title} {{Measurement of transverse momentum relative to dijet systems in PbPb and pp collisions at $ \sqrt{s_{\mathrm{NN}}}=2.76 $ TeV}},\ }\href {https://doi.org/10.1007/JHEP01(2016)006} {\bibfield  {journal} {\bibinfo  {journal} {JHEP}\ }\textbf {\bibinfo {volume} {01}},\ \bibinfo {pages} {006}},\ \Eprint {https://arxiv.org/abs/1509.09029} {arXiv:1509.09029 [nucl-ex]} \BibitemShut {NoStop}%
\bibitem [{\citenamefont {Sirunyan}\ \emph {et~al.}(2018)\citenamefont {Sirunyan} \emph {et~al.}}]{CMS:2017qlm}%
  \BibitemOpen
  \bibfield  {author} {\bibinfo {author} {\bibfnamefont {A.~M.}\ \bibnamefont {Sirunyan}} \emph {et~al.} (\bibinfo {collaboration} {CMS}),\ }\bibfield  {title} {\bibinfo {title} {{Measurement of the Splitting Function in $pp$ and Pb-Pb Collisions at $\sqrt{s_{_{\mathrm{NN}}}} =$ 5.02 TeV}},\ }\href {https://doi.org/10.1103/PhysRevLett.120.142302} {\bibfield  {journal} {\bibinfo  {journal} {Phys. Rev. Lett.}\ }\textbf {\bibinfo {volume} {120}},\ \bibinfo {pages} {142302} (\bibinfo {year} {2018})},\ \Eprint {https://arxiv.org/abs/1708.09429} {arXiv:1708.09429 [nucl-ex]} \BibitemShut {NoStop}%
\bibitem [{\citenamefont {Abdallah}\ \emph {et~al.}(2022)\citenamefont {Abdallah} \emph {et~al.}}]{STAR:2021kjt}%
  \BibitemOpen
  \bibfield  {author} {\bibinfo {author} {\bibfnamefont {M.~S.}\ \bibnamefont {Abdallah}} \emph {et~al.} (\bibinfo {collaboration} {STAR}),\ }\bibfield  {title} {\bibinfo {title} {{Differential measurements of jet substructure and partonic energy loss in Au+Au collisions at $\sqrt {S_{NN}}$ =200 GeV}},\ }\href {https://doi.org/10.1103/PhysRevC.105.044906} {\bibfield  {journal} {\bibinfo  {journal} {Phys. Rev. C}\ }\textbf {\bibinfo {volume} {105}},\ \bibinfo {pages} {044906} (\bibinfo {year} {2022})},\ \Eprint {https://arxiv.org/abs/2109.09793} {arXiv:2109.09793 [nucl-ex]} \BibitemShut {NoStop}%
\bibitem [{\citenamefont {Acharya}\ \emph {et~al.}(2022)\citenamefont {Acharya} \emph {et~al.}}]{ALargeIonColliderExperiment:2021mqf}%
  \BibitemOpen
  \bibfield  {author} {\bibinfo {author} {\bibfnamefont {S.}~\bibnamefont {Acharya}} \emph {et~al.} (\bibinfo {collaboration} {A Large Ion Collider Experiment, ALICE}),\ }\bibfield  {title} {\bibinfo {title} {{Measurement of the groomed jet radius and momentum splitting fraction in pp and Pb$-$Pb collisions at $\sqrt{s_{NN}} = 5.02$ TeV}},\ }\href {https://doi.org/10.1103/PhysRevLett.128.102001} {\bibfield  {journal} {\bibinfo  {journal} {Phys. Rev. Lett.}\ }\textbf {\bibinfo {volume} {128}},\ \bibinfo {pages} {102001} (\bibinfo {year} {2022})},\ \Eprint {https://arxiv.org/abs/2107.12984} {arXiv:2107.12984 [nucl-ex]} \BibitemShut {NoStop}%
\bibitem [{\citenamefont {Chatrchyan}\ \emph {et~al.}(2014{\natexlab{a}})\citenamefont {Chatrchyan} \emph {et~al.}}]{CMS:2013lhm}%
  \BibitemOpen
  \bibfield  {author} {\bibinfo {author} {\bibfnamefont {S.}~\bibnamefont {Chatrchyan}} \emph {et~al.} (\bibinfo {collaboration} {CMS}),\ }\bibfield  {title} {\bibinfo {title} {{Modification of Jet Shapes in PbPb Collisions at $\sqrt {s_{NN}} = 2.76$ TeV}},\ }\href {https://doi.org/10.1016/j.physletb.2014.01.042} {\bibfield  {journal} {\bibinfo  {journal} {Phys. Lett. B}\ }\textbf {\bibinfo {volume} {730}},\ \bibinfo {pages} {243} (\bibinfo {year} {2014}{\natexlab{a}})},\ \Eprint {https://arxiv.org/abs/1310.0878} {arXiv:1310.0878 [nucl-ex]} \BibitemShut {NoStop}%
\bibitem [{\citenamefont {Acharya}\ \emph {et~al.}(2019)\citenamefont {Acharya} \emph {et~al.}}]{ALICE:2019whv}%
  \BibitemOpen
  \bibfield  {author} {\bibinfo {author} {\bibfnamefont {S.}~\bibnamefont {Acharya}} \emph {et~al.} (\bibinfo {collaboration} {ALICE}),\ }\bibfield  {title} {\bibinfo {title} {{Measurement of jet radial profiles in Pb$–$Pb collisions at $\sqrt{s_{\rm NN}}=$ 2.76 TeV}},\ }\href {https://doi.org/10.1016/j.physletb.2019.07.020} {\bibfield  {journal} {\bibinfo  {journal} {Phys. Lett. B}\ }\textbf {\bibinfo {volume} {796}},\ \bibinfo {pages} {204} (\bibinfo {year} {2019})},\ \Eprint {https://arxiv.org/abs/1904.13118} {arXiv:1904.13118 [nucl-ex]} \BibitemShut {NoStop}%
\bibitem [{\citenamefont {Aad}\ \emph {et~al.}(2019)\citenamefont {Aad} \emph {et~al.}}]{ATLAS:2019pid}%
  \BibitemOpen
  \bibfield  {author} {\bibinfo {author} {\bibfnamefont {G.}~\bibnamefont {Aad}} \emph {et~al.} (\bibinfo {collaboration} {ATLAS}),\ }\bibfield  {title} {\bibinfo {title} {{Measurement of angular and momentum distributions of charged particles within and around jets in Pb+Pb and $pp$ collisions at $\sqrt{s_{\mathrm{NN}}} = 5.02$ TeV with the ATLAS detector}},\ }\href {https://doi.org/10.1103/PhysRevC.100.064901} {\bibfield  {journal} {\bibinfo  {journal} {Phys. Rev. C}\ }\textbf {\bibinfo {volume} {100}},\ \bibinfo {pages} {064901} (\bibinfo {year} {2019})},\ \bibinfo {note} {[Erratum: Phys.Rev.C 101, 059903 (2020)]},\ \Eprint {https://arxiv.org/abs/1908.05264} {arXiv:1908.05264 [nucl-ex]} \BibitemShut {NoStop}%
\bibitem [{\citenamefont {Chatrchyan}\ \emph {et~al.}(2014{\natexlab{b}})\citenamefont {Chatrchyan} \emph {et~al.}}]{CMS:2014jjt}%
  \BibitemOpen
  \bibfield  {author} {\bibinfo {author} {\bibfnamefont {S.}~\bibnamefont {Chatrchyan}} \emph {et~al.} (\bibinfo {collaboration} {CMS}),\ }\bibfield  {title} {\bibinfo {title} {{Measurement of Jet Fragmentation in PbPb and pp Collisions at $\sqrt{s_{NN}}= 2.76$ TeV}},\ }\href {https://doi.org/10.1103/PhysRevC.90.024908} {\bibfield  {journal} {\bibinfo  {journal} {Phys. Rev. C}\ }\textbf {\bibinfo {volume} {90}},\ \bibinfo {pages} {024908} (\bibinfo {year} {2014}{\natexlab{b}})},\ \Eprint {https://arxiv.org/abs/1406.0932} {arXiv:1406.0932 [nucl-ex]} \BibitemShut {NoStop}%
\bibitem [{\citenamefont {Aaboud}\ \emph {et~al.}(2017)\citenamefont {Aaboud} \emph {et~al.}}]{ATLAS:2017nre}%
  \BibitemOpen
  \bibfield  {author} {\bibinfo {author} {\bibfnamefont {M.}~\bibnamefont {Aaboud}} \emph {et~al.} (\bibinfo {collaboration} {ATLAS}),\ }\bibfield  {title} {\bibinfo {title} {{Measurement of jet fragmentation in Pb+Pb and $pp$ collisions at $\sqrt{{s_\mathrm{NN}}} = 2.76$ TeV with the ATLAS detector at the LHC}},\ }\href {https://doi.org/10.1140/epjc/s10052-017-4915-5} {\bibfield  {journal} {\bibinfo  {journal} {Eur. Phys. J. C}\ }\textbf {\bibinfo {volume} {77}},\ \bibinfo {pages} {379} (\bibinfo {year} {2017})},\ \Eprint {https://arxiv.org/abs/1702.00674} {arXiv:1702.00674 [hep-ex]} \BibitemShut {NoStop}%
\bibitem [{\citenamefont {Chen}\ \emph {et~al.}(2024)\citenamefont {Chen}, \citenamefont {Shen}, \citenamefont {Xue}, \citenamefont {Dai}, \citenamefont {Zhang},\ and\ \citenamefont {Wang}}]{Chen:2024cgx}%
  \BibitemOpen
  \bibfield  {author} {\bibinfo {author} {\bibfnamefont {S.-Y.}\ \bibnamefont {Chen}}, \bibinfo {author} {\bibfnamefont {K.-M.}\ \bibnamefont {Shen}}, \bibinfo {author} {\bibfnamefont {X.-F.}\ \bibnamefont {Xue}}, \bibinfo {author} {\bibfnamefont {W.}~\bibnamefont {Dai}}, \bibinfo {author} {\bibfnamefont {B.-W.}\ \bibnamefont {Zhang}},\ and\ \bibinfo {author} {\bibfnamefont {E.-K.}\ \bibnamefont {Wang}},\ }\href {https://arxiv.org/abs/2409.13996} {\bibinfo {title} {Study of eec discrimination power on quark and gluon quenching effects in heavy-ion collisions at $\sqrt{s}=5.02$ tev}} (\bibinfo {year} {2024}),\ \Eprint {https://arxiv.org/abs/2409.13996} {arXiv:2409.13996 [nucl-th]} \BibitemShut {NoStop}%
\bibitem [{\citenamefont {{HEP ML Community}}()}]{hepmllivingreview}%
  \BibitemOpen
  \bibfield  {author} {\bibinfo {author} {\bibnamefont {{HEP ML Community}}},\ }\href {https://iml-wg.github.io/HEPML-LivingReview/} {\bibinfo {title} {{A Living Review of Machine Learning for Particle Physics}}}\BibitemShut {NoStop}%
\bibitem [{\citenamefont {Liu}\ \emph {et~al.}(2023)\citenamefont {Liu}, \citenamefont {Velkovska}, \citenamefont {Wu},\ and\ \citenamefont {Verweij}}]{Liu:2022hzd}%
  \BibitemOpen
  \bibfield  {author} {\bibinfo {author} {\bibfnamefont {L.}~\bibnamefont {Liu}}, \bibinfo {author} {\bibfnamefont {J.}~\bibnamefont {Velkovska}}, \bibinfo {author} {\bibfnamefont {Y.}~\bibnamefont {Wu}},\ and\ \bibinfo {author} {\bibfnamefont {M.}~\bibnamefont {Verweij}},\ }\bibfield  {title} {\bibinfo {title} {{Identifying quenched jets in heavy ion collisions with machine learning}},\ }\href {https://doi.org/10.1007/JHEP04(2023)140} {\bibfield  {journal} {\bibinfo  {journal} {JHEP}\ }\textbf {\bibinfo {volume} {04}},\ \bibinfo {pages} {140}},\ \Eprint {https://arxiv.org/abs/2206.01628} {arXiv:2206.01628 [hep-ph]} \BibitemShut {NoStop}%
\bibitem [{\citenamefont {Apolin\'ario}\ \emph {et~al.}(2021)\citenamefont {Apolin\'ario}, \citenamefont {Castro}, \citenamefont {Crispim Rom\~ao}, \citenamefont {Milhano}, \citenamefont {Pedro},\ and\ \citenamefont {Peres}}]{Apolinario:2021olp}%
  \BibitemOpen
  \bibfield  {author} {\bibinfo {author} {\bibfnamefont {L.}~\bibnamefont {Apolin\'ario}}, \bibinfo {author} {\bibfnamefont {N.~F.}\ \bibnamefont {Castro}}, \bibinfo {author} {\bibfnamefont {M.}~\bibnamefont {Crispim Rom\~ao}}, \bibinfo {author} {\bibfnamefont {J.~G.}\ \bibnamefont {Milhano}}, \bibinfo {author} {\bibfnamefont {R.}~\bibnamefont {Pedro}},\ and\ \bibinfo {author} {\bibfnamefont {F.~C.~R.}\ \bibnamefont {Peres}},\ }\bibfield  {title} {\bibinfo {title} {{Deep Learning for the classification of quenched jets}},\ }\href {https://doi.org/10.1007/JHEP11(2021)219} {\bibfield  {journal} {\bibinfo  {journal} {JHEP}\ }\textbf {\bibinfo {volume} {11}},\ \bibinfo {pages} {219}},\ \Eprint {https://arxiv.org/abs/2106.08869} {arXiv:2106.08869 [hep-ph]} \BibitemShut {NoStop}%
\bibitem [{\citenamefont {Lai}\ \emph {et~al.}(2022)\citenamefont {Lai}, \citenamefont {Mulligan}, \citenamefont {P\l{}osko\'n},\ and\ \citenamefont {Ringer}}]{Lai:2021ckt}%
  \BibitemOpen
  \bibfield  {author} {\bibinfo {author} {\bibfnamefont {Y.~S.}\ \bibnamefont {Lai}}, \bibinfo {author} {\bibfnamefont {J.}~\bibnamefont {Mulligan}}, \bibinfo {author} {\bibfnamefont {M.}~\bibnamefont {P\l{}osko\'n}},\ and\ \bibinfo {author} {\bibfnamefont {F.}~\bibnamefont {Ringer}},\ }\bibfield  {title} {\bibinfo {title} {{The information content of jet quenching and machine learning assisted observable design}},\ }\href {https://doi.org/10.1007/JHEP10(2022)011} {\bibfield  {journal} {\bibinfo  {journal} {JHEP}\ }\textbf {\bibinfo {volume} {10}},\ \bibinfo {pages} {011}},\ \Eprint {https://arxiv.org/abs/2111.14589} {arXiv:2111.14589 [hep-ph]} \BibitemShut {NoStop}%
\bibitem [{\citenamefont {Crispim Rom\~ao}\ \emph {et~al.}(2024)\citenamefont {Crispim Rom\~ao}, \citenamefont {Milhano},\ and\ \citenamefont {van Leeuwen}}]{CrispimRomao:2023ssj}%
  \BibitemOpen
  \bibfield  {author} {\bibinfo {author} {\bibfnamefont {M.}~\bibnamefont {Crispim Rom\~ao}}, \bibinfo {author} {\bibfnamefont {J.~G.}\ \bibnamefont {Milhano}},\ and\ \bibinfo {author} {\bibfnamefont {M.}~\bibnamefont {van Leeuwen}},\ }\bibfield  {title} {\bibinfo {title} {{Jet substructure observables for jet quenching in quark gluon plasma: A machine learning driven analysis}},\ }\href {https://doi.org/10.21468/SciPostPhys.16.1.015} {\bibfield  {journal} {\bibinfo  {journal} {SciPost Phys.}\ }\textbf {\bibinfo {volume} {16}},\ \bibinfo {pages} {015} (\bibinfo {year} {2024})},\ \Eprint {https://arxiv.org/abs/2304.07196} {arXiv:2304.07196 [hep-ph]} \BibitemShut {NoStop}%
\bibitem [{\citenamefont {Du}(2023)}]{Du:2023qst}%
  \BibitemOpen
  \bibfield  {author} {\bibinfo {author} {\bibfnamefont {Y.-L.}\ \bibnamefont {Du}},\ }\bibfield  {title} {\bibinfo {title} {{Overview: Jet quenching with machine learning}},\ }in\ \href@noop {} {\emph {\bibinfo {booktitle} {{11th International Conference on Hard and Electromagnetic Probes of High-Energy Nuclear Collisions}: {Hard Probes 2023}}}}\ (\bibinfo {year} {2023})\ \Eprint {https://arxiv.org/abs/2308.10035} {arXiv:2308.10035 [hep-ph]} \BibitemShut {NoStop}%
\bibitem [{\citenamefont {Zapp}\ \emph {et~al.}(2013)\citenamefont {Zapp}, \citenamefont {Krauss},\ and\ \citenamefont {Wiedemann}}]{Zapp:2012ak}%
  \BibitemOpen
  \bibfield  {author} {\bibinfo {author} {\bibfnamefont {K.~C.}\ \bibnamefont {Zapp}}, \bibinfo {author} {\bibfnamefont {F.}~\bibnamefont {Krauss}},\ and\ \bibinfo {author} {\bibfnamefont {U.~A.}\ \bibnamefont {Wiedemann}},\ }\bibfield  {title} {\bibinfo {title} {{A perturbative framework for jet quenching}},\ }\href {https://doi.org/10.1007/JHEP03(2013)080} {\bibfield  {journal} {\bibinfo  {journal} {JHEP}\ }\textbf {\bibinfo {volume} {03}},\ \bibinfo {pages} {080}},\ \Eprint {https://arxiv.org/abs/1212.1599} {arXiv:1212.1599 [hep-ph]} \BibitemShut {NoStop}%
\bibitem [{\citenamefont {Chen}\ \emph {et~al.}(2018)\citenamefont {Chen}, \citenamefont {Cao}, \citenamefont {Luo}, \citenamefont {Pang},\ and\ \citenamefont {Wang}}]{Chen:2017zte}%
  \BibitemOpen
  \bibfield  {author} {\bibinfo {author} {\bibfnamefont {W.}~\bibnamefont {Chen}}, \bibinfo {author} {\bibfnamefont {S.}~\bibnamefont {Cao}}, \bibinfo {author} {\bibfnamefont {T.}~\bibnamefont {Luo}}, \bibinfo {author} {\bibfnamefont {L.-G.}\ \bibnamefont {Pang}},\ and\ \bibinfo {author} {\bibfnamefont {X.-N.}\ \bibnamefont {Wang}},\ }\bibfield  {title} {\bibinfo {title} {{Effects of jet-induced medium excitation in $\gamma$-hadron correlation in A+A collisions}},\ }\href {https://doi.org/10.1016/j.physletb.2017.12.015} {\bibfield  {journal} {\bibinfo  {journal} {Phys. Lett. B}\ }\textbf {\bibinfo {volume} {777}},\ \bibinfo {pages} {86} (\bibinfo {year} {2018})},\ \Eprint {https://arxiv.org/abs/1704.03648} {arXiv:1704.03648 [nucl-th]} \BibitemShut {NoStop}%
\bibitem [{\citenamefont {Sj\"ostrand}\ \emph {et~al.}(2015)\citenamefont {Sj\"ostrand}, \citenamefont {Ask}, \citenamefont {Christiansen}, \citenamefont {Corke}, \citenamefont {Desai}, \citenamefont {Ilten}, \citenamefont {Mrenna}, \citenamefont {Prestel}, \citenamefont {Rasmussen},\ and\ \citenamefont {Skands}}]{Sjostrand:2014zea}%
  \BibitemOpen
  \bibfield  {author} {\bibinfo {author} {\bibfnamefont {T.}~\bibnamefont {Sj\"ostrand}}, \bibinfo {author} {\bibfnamefont {S.}~\bibnamefont {Ask}}, \bibinfo {author} {\bibfnamefont {J.~R.}\ \bibnamefont {Christiansen}}, \bibinfo {author} {\bibfnamefont {R.}~\bibnamefont {Corke}}, \bibinfo {author} {\bibfnamefont {N.}~\bibnamefont {Desai}}, \bibinfo {author} {\bibfnamefont {P.}~\bibnamefont {Ilten}}, \bibinfo {author} {\bibfnamefont {S.}~\bibnamefont {Mrenna}}, \bibinfo {author} {\bibfnamefont {S.}~\bibnamefont {Prestel}}, \bibinfo {author} {\bibfnamefont {C.~O.}\ \bibnamefont {Rasmussen}},\ and\ \bibinfo {author} {\bibfnamefont {P.~Z.}\ \bibnamefont {Skands}},\ }\bibfield  {title} {\bibinfo {title} {{An introduction to PYTHIA 8.2}},\ }\href {https://doi.org/10.1016/j.cpc.2015.01.024} {\bibfield  {journal} {\bibinfo  {journal} {Comput. Phys. Commun.}\ }\textbf {\bibinfo {volume} {191}},\ \bibinfo {pages} {159} (\bibinfo {year} {2015})},\ \Eprint {https://arxiv.org/abs/1410.3012} {arXiv:1410.3012 [hep-ph]}
  \BibitemShut {NoStop}%
\bibitem [{\citenamefont {Sjostrand}\ \emph {et~al.}(2008)\citenamefont {Sjostrand}, \citenamefont {Mrenna},\ and\ \citenamefont {Skands}}]{Sjostrand:2007gs}%
  \BibitemOpen
  \bibfield  {author} {\bibinfo {author} {\bibfnamefont {T.}~\bibnamefont {Sjostrand}}, \bibinfo {author} {\bibfnamefont {S.}~\bibnamefont {Mrenna}},\ and\ \bibinfo {author} {\bibfnamefont {P.~Z.}\ \bibnamefont {Skands}},\ }\bibfield  {title} {\bibinfo {title} {{A Brief Introduction to PYTHIA 8.1}},\ }\href {https://doi.org/10.1016/j.cpc.2008.01.036} {\bibfield  {journal} {\bibinfo  {journal} {Comput. Phys. Commun.}\ }\textbf {\bibinfo {volume} {178}},\ \bibinfo {pages} {852} (\bibinfo {year} {2008})},\ \Eprint {https://arxiv.org/abs/0710.3820} {arXiv:0710.3820 [hep-ph]} \BibitemShut {NoStop}%
\bibitem [{\citenamefont {Bewick}\ \emph {et~al.}(2024)\citenamefont {Bewick}, \citenamefont {Ravasio}, \citenamefont {Gieseke}, \citenamefont {Kiebacher}, \citenamefont {Masouminia}, \citenamefont {Papaefstathiou}, \citenamefont {Plätzer}, \citenamefont {Richardson}, \citenamefont {Samitz}, \citenamefont {Seymour}, \citenamefont {Siódmok},\ and\ \citenamefont {Whitehead}}]{Bewick:2023tfi}%
  \BibitemOpen
  \bibfield  {author} {\bibinfo {author} {\bibfnamefont {G.}~\bibnamefont {Bewick}}, \bibinfo {author} {\bibfnamefont {S.~F.}\ \bibnamefont {Ravasio}}, \bibinfo {author} {\bibfnamefont {S.}~\bibnamefont {Gieseke}}, \bibinfo {author} {\bibfnamefont {S.}~\bibnamefont {Kiebacher}}, \bibinfo {author} {\bibfnamefont {M.~R.}\ \bibnamefont {Masouminia}}, \bibinfo {author} {\bibfnamefont {A.}~\bibnamefont {Papaefstathiou}}, \bibinfo {author} {\bibfnamefont {S.}~\bibnamefont {Plätzer}}, \bibinfo {author} {\bibfnamefont {P.}~\bibnamefont {Richardson}}, \bibinfo {author} {\bibfnamefont {D.}~\bibnamefont {Samitz}}, \bibinfo {author} {\bibfnamefont {M.~H.}\ \bibnamefont {Seymour}}, \bibinfo {author} {\bibfnamefont {A.}~\bibnamefont {Siódmok}},\ and\ \bibinfo {author} {\bibfnamefont {J.}~\bibnamefont {Whitehead}},\ }\href {https://arxiv.org/abs/2312.05175} {\bibinfo {title} {Herwig 7.3 release note}} (\bibinfo {year} {2024}),\ \Eprint {https://arxiv.org/abs/2312.05175} {arXiv:2312.05175 [hep-ph]} \BibitemShut {NoStop}%
\bibitem [{\citenamefont {Bahr}\ \emph {et~al.}(2008)\citenamefont {Bahr} \emph {et~al.}}]{Bahr:2008pv}%
  \BibitemOpen
  \bibfield  {author} {\bibinfo {author} {\bibfnamefont {M.}~\bibnamefont {Bahr}} \emph {et~al.},\ }\bibfield  {title} {\bibinfo {title} {{Herwig++ Physics and Manual}},\ }\href {https://doi.org/10.1140/epjc/s10052-008-0798-9} {\bibfield  {journal} {\bibinfo  {journal} {Eur. Phys. J. C}\ }\textbf {\bibinfo {volume} {58}},\ \bibinfo {pages} {639} (\bibinfo {year} {2008})},\ \Eprint {https://arxiv.org/abs/0803.0883} {arXiv:0803.0883 [hep-ph]} \BibitemShut {NoStop}%
\bibitem [{\citenamefont {Skands}\ \emph {et~al.}(2014)\citenamefont {Skands}, \citenamefont {Carrazza},\ and\ \citenamefont {Rojo}}]{Skands:2014pea}%
  \BibitemOpen
  \bibfield  {author} {\bibinfo {author} {\bibfnamefont {P.}~\bibnamefont {Skands}}, \bibinfo {author} {\bibfnamefont {S.}~\bibnamefont {Carrazza}},\ and\ \bibinfo {author} {\bibfnamefont {J.}~\bibnamefont {Rojo}},\ }\bibfield  {title} {\bibinfo {title} {{Tuning PYTHIA 8.1: the Monash 2013 Tune}},\ }\href {https://doi.org/10.1140/epjc/s10052-014-3024-y} {\bibfield  {journal} {\bibinfo  {journal} {Eur. Phys. J. C}\ }\textbf {\bibinfo {volume} {74}},\ \bibinfo {pages} {3024} (\bibinfo {year} {2014})},\ \Eprint {https://arxiv.org/abs/1404.5630} {arXiv:1404.5630 [hep-ph]} \BibitemShut {NoStop}%
\bibitem [{\citenamefont {Qureshi}\ \emph {et~al.}(2024{\natexlab{a}})\citenamefont {Qureshi}, \citenamefont {Elayavalli}, \citenamefont {Mozarsky}, \citenamefont {Caines},\ and\ \citenamefont {Mooney}}]{qureshi2024newherwig7underlyingevent}%
  \BibitemOpen
  \bibfield  {author} {\bibinfo {author} {\bibfnamefont {U.~S.}\ \bibnamefont {Qureshi}}, \bibinfo {author} {\bibfnamefont {R.~K.}\ \bibnamefont {Elayavalli}}, \bibinfo {author} {\bibfnamefont {L.}~\bibnamefont {Mozarsky}}, \bibinfo {author} {\bibfnamefont {H.}~\bibnamefont {Caines}},\ and\ \bibinfo {author} {\bibfnamefont {I.}~\bibnamefont {Mooney}},\ }\href {https://arxiv.org/abs/2411.16897} {\bibinfo {title} {A new herwig7 underlying event tune: from rhic to lhc energies}} (\bibinfo {year} {2024}{\natexlab{a}}),\ \Eprint {https://arxiv.org/abs/2411.16897} {arXiv:2411.16897 [hep-ph]} \BibitemShut {NoStop}%
\bibitem [{\citenamefont {Andrews}\ \emph {et~al.}(2020)\citenamefont {Andrews} \emph {et~al.}}]{Andrews:2018jcm}%
  \BibitemOpen
  \bibfield  {author} {\bibinfo {author} {\bibfnamefont {H.~A.}\ \bibnamefont {Andrews}} \emph {et~al.},\ }\bibfield  {title} {\bibinfo {title} {{Novel tools and observables for jet physics in heavy-ion collisions}},\ }\href {https://doi.org/10.1088/1361-6471/ab7cbc} {\bibfield  {journal} {\bibinfo  {journal} {J. Phys. G}\ }\textbf {\bibinfo {volume} {47}},\ \bibinfo {pages} {065102} (\bibinfo {year} {2020})},\ \Eprint {https://arxiv.org/abs/1808.03689} {arXiv:1808.03689 [hep-ph]} \BibitemShut {NoStop}%
\bibitem [{\citenamefont {Adam}\ \emph {et~al.}(2017)\citenamefont {Adam} \emph {et~al.}}]{ALICE:2016fbt}%
  \BibitemOpen
  \bibfield  {author} {\bibinfo {author} {\bibfnamefont {J.}~\bibnamefont {Adam}} \emph {et~al.} (\bibinfo {collaboration} {ALICE}),\ }\bibfield  {title} {\bibinfo {title} {{Centrality dependence of the pseudorapidity density distribution for charged particles in Pb-Pb collisions at $\sqrt{s_{\rm NN}}=5.02$ TeV}},\ }\href {https://doi.org/10.1016/j.physletb.2017.07.017} {\bibfield  {journal} {\bibinfo  {journal} {Phys. Lett. B}\ }\textbf {\bibinfo {volume} {772}},\ \bibinfo {pages} {567} (\bibinfo {year} {2017})},\ \Eprint {https://arxiv.org/abs/1612.08966} {arXiv:1612.08966 [nucl-ex]} \BibitemShut {NoStop}%
\bibitem [{\citenamefont {Berta}\ \emph {et~al.}(2014)\citenamefont {Berta}, \citenamefont {Spousta}, \citenamefont {Miller},\ and\ \citenamefont {Leitner}}]{Berta:2014eza}%
  \BibitemOpen
  \bibfield  {author} {\bibinfo {author} {\bibfnamefont {P.}~\bibnamefont {Berta}}, \bibinfo {author} {\bibfnamefont {M.}~\bibnamefont {Spousta}}, \bibinfo {author} {\bibfnamefont {D.~W.}\ \bibnamefont {Miller}},\ and\ \bibinfo {author} {\bibfnamefont {R.}~\bibnamefont {Leitner}},\ }\bibfield  {title} {\bibinfo {title} {{Particle-level pileup subtraction for jets and jet shapes}},\ }\href {https://doi.org/10.1007/JHEP06(2014)092} {\bibfield  {journal} {\bibinfo  {journal} {JHEP}\ }\textbf {\bibinfo {volume} {06}},\ \bibinfo {pages} {092}},\ \Eprint {https://arxiv.org/abs/1403.3108} {arXiv:1403.3108 [hep-ex]} \BibitemShut {NoStop}%
\bibitem [{\citenamefont {Cacciari}\ \emph {et~al.}(2008)\citenamefont {Cacciari}, \citenamefont {Salam},\ and\ \citenamefont {Soyez}}]{Cacciari:2008gp}%
  \BibitemOpen
  \bibfield  {author} {\bibinfo {author} {\bibfnamefont {M.}~\bibnamefont {Cacciari}}, \bibinfo {author} {\bibfnamefont {G.~P.}\ \bibnamefont {Salam}},\ and\ \bibinfo {author} {\bibfnamefont {G.}~\bibnamefont {Soyez}},\ }\bibfield  {title} {\bibinfo {title} {{The anti-$k_t$ jet clustering algorithm}},\ }\href {https://doi.org/10.1088/1126-6708/2008/04/063} {\bibfield  {journal} {\bibinfo  {journal} {JHEP}\ }\textbf {\bibinfo {volume} {04}},\ \bibinfo {pages} {063}},\ \Eprint {https://arxiv.org/abs/0802.1189} {arXiv:0802.1189 [hep-ph]} \BibitemShut {NoStop}%
\bibitem [{\citenamefont {Cacciari}\ \emph {et~al.}(2012)\citenamefont {Cacciari}, \citenamefont {Salam},\ and\ \citenamefont {Soyez}}]{Cacciari:2011ma}%
  \BibitemOpen
  \bibfield  {author} {\bibinfo {author} {\bibfnamefont {M.}~\bibnamefont {Cacciari}}, \bibinfo {author} {\bibfnamefont {G.~P.}\ \bibnamefont {Salam}},\ and\ \bibinfo {author} {\bibfnamefont {G.}~\bibnamefont {Soyez}},\ }\bibfield  {title} {\bibinfo {title} {{FastJet User Manual}},\ }\href {https://doi.org/10.1140/epjc/s10052-012-1896-2} {\bibfield  {journal} {\bibinfo  {journal} {Eur. Phys. J. C}\ }\textbf {\bibinfo {volume} {72}},\ \bibinfo {pages} {1896} (\bibinfo {year} {2012})},\ \Eprint {https://arxiv.org/abs/1111.6097} {arXiv:1111.6097 [hep-ph]} \BibitemShut {NoStop}%
\bibitem [{\citenamefont {Cacciari}\ and\ \citenamefont {Salam}(2006)}]{Cacciari:2005hq}%
  \BibitemOpen
  \bibfield  {author} {\bibinfo {author} {\bibfnamefont {M.}~\bibnamefont {Cacciari}}\ and\ \bibinfo {author} {\bibfnamefont {G.~P.}\ \bibnamefont {Salam}},\ }\bibfield  {title} {\bibinfo {title} {{Dispelling the $N^{3}$ myth for the $k_t$ jet-finder}},\ }\href {https://doi.org/10.1016/j.physletb.2006.08.037} {\bibfield  {journal} {\bibinfo  {journal} {Phys. Lett. B}\ }\textbf {\bibinfo {volume} {641}},\ \bibinfo {pages} {57} (\bibinfo {year} {2006})},\ \Eprint {https://arxiv.org/abs/hep-ph/0512210} {arXiv:hep-ph/0512210} \BibitemShut {NoStop}%
\bibitem [{\citenamefont {Larkoski}\ \emph {et~al.}(2014)\citenamefont {Larkoski}, \citenamefont {Marzani}, \citenamefont {Soyez},\ and\ \citenamefont {Thaler}}]{Larkoski:2014wba}%
  \BibitemOpen
  \bibfield  {author} {\bibinfo {author} {\bibfnamefont {A.~J.}\ \bibnamefont {Larkoski}}, \bibinfo {author} {\bibfnamefont {S.}~\bibnamefont {Marzani}}, \bibinfo {author} {\bibfnamefont {G.}~\bibnamefont {Soyez}},\ and\ \bibinfo {author} {\bibfnamefont {J.}~\bibnamefont {Thaler}},\ }\bibfield  {title} {\bibinfo {title} {{Soft Drop}},\ }\href {https://doi.org/10.1007/JHEP05(2014)146} {\bibfield  {journal} {\bibinfo  {journal} {JHEP}\ }\textbf {\bibinfo {volume} {05}},\ \bibinfo {pages} {146}},\ \Eprint {https://arxiv.org/abs/1402.2657} {arXiv:1402.2657 [hep-ph]} \BibitemShut {NoStop}%
\bibitem [{\citenamefont {Arik}\ and\ \citenamefont {Pfister}(2020)}]{arik2020tabnetattentiveinterpretabletabular}%
  \BibitemOpen
  \bibfield  {author} {\bibinfo {author} {\bibfnamefont {S.~O.}\ \bibnamefont {Arik}}\ and\ \bibinfo {author} {\bibfnamefont {T.}~\bibnamefont {Pfister}},\ }\href {https://arxiv.org/abs/1908.07442} {\bibinfo {title} {Tabnet: Attentive interpretable tabular learning}} (\bibinfo {year} {2020}),\ \Eprint {https://arxiv.org/abs/1908.07442} {arXiv:1908.07442 [cs.LG]} \BibitemShut {NoStop}%
\bibitem [{\citenamefont {Ai}\ \emph {et~al.}(2023)\citenamefont {Ai}, \citenamefont {Hsu}, \citenamefont {Li},\ and\ \citenamefont {Lu}}]{Ai:2022qvs}%
  \BibitemOpen
  \bibfield  {author} {\bibinfo {author} {\bibfnamefont {X.}~\bibnamefont {Ai}}, \bibinfo {author} {\bibfnamefont {S.~C.}\ \bibnamefont {Hsu}}, \bibinfo {author} {\bibfnamefont {K.}~\bibnamefont {Li}},\ and\ \bibinfo {author} {\bibfnamefont {C.~T.}\ \bibnamefont {Lu}},\ }\bibfield  {title} {\bibinfo {title} {Probing highly collimated photon-jets with deep learning},\ }\href {https://doi.org/10.1088/1742-6596/2438/1/012114} {\bibfield  {journal} {\bibinfo  {journal} {JPCS}\ }\textbf {\bibinfo {volume} {2438}},\ \bibinfo {pages} {012114} (\bibinfo {year} {2023})}\BibitemShut {NoStop}%
\bibitem [{\citenamefont {Collaboration}(2018)}]{ATLAS:2017fak}%
  \BibitemOpen
  \bibfield  {author} {\bibinfo {author} {\bibfnamefont {A.}~\bibnamefont {Collaboration}} (\bibinfo {collaboration} {ATLAS}),\ }\bibfield  {title} {\bibinfo {title} {{Search for the standard model Higgs boson produced in association with top quarks and decaying into a $b\bar{b}$ pair in $pp$ collisions at $\sqrt{s}$ = 13 TeV with the ATLAS detector}},\ }\href {https://doi.org/10.1103/PhysRevD.97.072016} {\bibfield  {journal} {\bibinfo  {journal} {Phys. Rev. D}\ }\textbf {\bibinfo {volume} {97}},\ \bibinfo {pages} {072016} (\bibinfo {year} {2018})},\ \Eprint {https://arxiv.org/abs/1712.08895} {arXiv:1712.08895 [hep-ex]} \BibitemShut {NoStop}%
\bibitem [{\citenamefont {Arganda}\ \emph {et~al.}(2024)\citenamefont {Arganda}, \citenamefont {D\'{\i}az}, \citenamefont {Perez}, \citenamefont {Sand\'a~Seoane},\ and\ \citenamefont {Szynkman}}]{Arganda2024}%
  \BibitemOpen
  \bibfield  {author} {\bibinfo {author} {\bibfnamefont {E.}~\bibnamefont {Arganda}}, \bibinfo {author} {\bibfnamefont {D.~A.}\ \bibnamefont {D\'{\i}az}}, \bibinfo {author} {\bibfnamefont {A.~D.}\ \bibnamefont {Perez}}, \bibinfo {author} {\bibfnamefont {R.~M.}\ \bibnamefont {Sand\'a~Seoane}},\ and\ \bibinfo {author} {\bibfnamefont {A.}~\bibnamefont {Szynkman}},\ }\bibfield  {title} {\bibinfo {title} {Lhc study of third-generation scalar leptoquarks with machine-learned likelihoods},\ }\href {https://doi.org/10.1103/PhysRevD.109.055032} {\bibfield  {journal} {\bibinfo  {journal} {Phys. Rev. D}\ }\textbf {\bibinfo {volume} {109}},\ \bibinfo {pages} {055032} (\bibinfo {year} {2024})}\BibitemShut {NoStop}%
\bibitem [{\citenamefont {Flórez}\ \emph {et~al.}(2024)\citenamefont {Flórez}, \citenamefont {Gurrola}, \citenamefont {Rodriguez},\ and\ \citenamefont {Qureshi}}]{flórez2024probinglightscalarsvectorlike}%
  \BibitemOpen
  \bibfield  {author} {\bibinfo {author} {\bibfnamefont {A.}~\bibnamefont {Flórez}}, \bibinfo {author} {\bibfnamefont {A.}~\bibnamefont {Gurrola}}, \bibinfo {author} {\bibfnamefont {C.}~\bibnamefont {Rodriguez}},\ and\ \bibinfo {author} {\bibfnamefont {U.~S.}\ \bibnamefont {Qureshi}},\ }\href {https://arxiv.org/abs/2410.17854} {\bibinfo {title} {Probing light scalars and vector-like quarks at the high-luminosity lhc}} (\bibinfo {year} {2024}),\ \Eprint {https://arxiv.org/abs/2410.17854} {arXiv:2410.17854 [hep-ph]} \BibitemShut {NoStop}%
\bibitem [{\citenamefont {Whiteson}(2014)}]{misc_higgs_280}%
  \BibitemOpen
  \bibfield  {author} {\bibinfo {author} {\bibfnamefont {D.}~\bibnamefont {Whiteson}},\ }\href@noop {} {\bibinfo {title} {{HIGGS}}},\ \bibinfo {howpublished} {UCI Machine Learning Repository} (\bibinfo {year} {2014}),\ \bibinfo {note} {{DOI}: https://doi.org/10.24432/C5V312}\BibitemShut {NoStop}%
\bibitem [{\citenamefont {Qureshi}\ \emph {et~al.}(2024{\natexlab{b}})\citenamefont {Qureshi}, \citenamefont {Gurrola},\ and\ \citenamefont {Flórez}}]{qureshi2024probingcompressedmassspectrum}%
  \BibitemOpen
  \bibfield  {author} {\bibinfo {author} {\bibfnamefont {U.~S.}\ \bibnamefont {Qureshi}}, \bibinfo {author} {\bibfnamefont {A.}~\bibnamefont {Gurrola}},\ and\ \bibinfo {author} {\bibfnamefont {A.}~\bibnamefont {Flórez}},\ }\href {https://arxiv.org/abs/2411.13837} {\bibinfo {title} {Probing compressed mass spectrum supersymmetry at the lhc with the vector boson fusion topology}} (\bibinfo {year} {2024}{\natexlab{b}}),\ \Eprint {https://arxiv.org/abs/2411.13837} {arXiv:2411.13837 [hep-ph]} \BibitemShut {NoStop}%
\bibitem [{\citenamefont {Mikuni}\ and\ \citenamefont {Canelli}(2020)}]{ABCNet}%
  \BibitemOpen
  \bibfield  {author} {\bibinfo {author} {\bibfnamefont {V.}~\bibnamefont {Mikuni}}\ and\ \bibinfo {author} {\bibfnamefont {F.}~\bibnamefont {Canelli}},\ }\bibfield  {title} {\bibinfo {title} {Abcnet: an attention-based method for particle tagging},\ }\bibfield  {journal} {\bibinfo  {journal} {The European Physical Journal Plus}\ }\textbf {\bibinfo {volume} {135}},\ \href {https://doi.org/10.1140/epjp/s13360-020-00497-3} {10.1140/epjp/s13360-020-00497-3} (\bibinfo {year} {2020})\BibitemShut {NoStop}%
\bibitem [{\citenamefont {Paszke}\ \emph {et~al.}(2019)\citenamefont {Paszke}, \citenamefont {Gross}, \citenamefont {Massa}, \citenamefont {Lerer}, \citenamefont {Bradbury}, \citenamefont {Chanan}, \citenamefont {Killeen}, \citenamefont {Lin}, \citenamefont {Gimelshein}, \citenamefont {Antiga}, \citenamefont {Desmaison}, \citenamefont {Köpf}, \citenamefont {Yang}, \citenamefont {DeVito}, \citenamefont {Raison}, \citenamefont {Tejani}, \citenamefont {Chilamkurthy}, \citenamefont {Steiner}, \citenamefont {Fang}, \citenamefont {Bai},\ and\ \citenamefont {Chintala}}]{paszke2019pytorchimperativestylehighperformance}%
  \BibitemOpen
  \bibfield  {author} {\bibinfo {author} {\bibfnamefont {A.}~\bibnamefont {Paszke}}, \bibinfo {author} {\bibfnamefont {S.}~\bibnamefont {Gross}}, \bibinfo {author} {\bibfnamefont {F.}~\bibnamefont {Massa}}, \bibinfo {author} {\bibfnamefont {A.}~\bibnamefont {Lerer}}, \bibinfo {author} {\bibfnamefont {J.}~\bibnamefont {Bradbury}}, \bibinfo {author} {\bibfnamefont {G.}~\bibnamefont {Chanan}}, \bibinfo {author} {\bibfnamefont {T.}~\bibnamefont {Killeen}}, \bibinfo {author} {\bibfnamefont {Z.}~\bibnamefont {Lin}}, \bibinfo {author} {\bibfnamefont {N.}~\bibnamefont {Gimelshein}}, \bibinfo {author} {\bibfnamefont {L.}~\bibnamefont {Antiga}}, \bibinfo {author} {\bibfnamefont {A.}~\bibnamefont {Desmaison}}, \bibinfo {author} {\bibfnamefont {A.}~\bibnamefont {Köpf}}, \bibinfo {author} {\bibfnamefont {E.}~\bibnamefont {Yang}}, \bibinfo {author} {\bibfnamefont {Z.}~\bibnamefont {DeVito}}, \bibinfo {author} {\bibfnamefont {M.}~\bibnamefont {Raison}}, \bibinfo {author} {\bibfnamefont {A.}~\bibnamefont {Tejani}}, \bibinfo
  {author} {\bibfnamefont {S.}~\bibnamefont {Chilamkurthy}}, \bibinfo {author} {\bibfnamefont {B.}~\bibnamefont {Steiner}}, \bibinfo {author} {\bibfnamefont {L.}~\bibnamefont {Fang}}, \bibinfo {author} {\bibfnamefont {J.}~\bibnamefont {Bai}},\ and\ \bibinfo {author} {\bibfnamefont {S.}~\bibnamefont {Chintala}},\ }\href {https://arxiv.org/abs/1912.01703} {\bibinfo {title} {Pytorch: An imperative style, high-performance deep learning library}} (\bibinfo {year} {2019}),\ \Eprint {https://arxiv.org/abs/1912.01703} {arXiv:1912.01703 [cs.LG]} \BibitemShut {NoStop}%
\bibitem [{\citenamefont {Pedregosa}\ \emph {et~al.}(2011)\citenamefont {Pedregosa}, \citenamefont {Varoquaux}, \citenamefont {Gramfort}, \citenamefont {Michel}, \citenamefont {Thirion}, \citenamefont {Grisel}, \citenamefont {Blondel}, \citenamefont {Prettenhofer}, \citenamefont {Weiss}, \citenamefont {Dubourg}, \citenamefont {Vanderplas}, \citenamefont {Passos}, \citenamefont {Cournapeau}, \citenamefont {Brucher}, \citenamefont {Perrot},\ and\ \citenamefont {Duchesnay}}]{scikit-learn}%
  \BibitemOpen
  \bibfield  {author} {\bibinfo {author} {\bibfnamefont {F.}~\bibnamefont {Pedregosa}}, \bibinfo {author} {\bibfnamefont {G.}~\bibnamefont {Varoquaux}}, \bibinfo {author} {\bibfnamefont {A.}~\bibnamefont {Gramfort}}, \bibinfo {author} {\bibfnamefont {V.}~\bibnamefont {Michel}}, \bibinfo {author} {\bibfnamefont {B.}~\bibnamefont {Thirion}}, \bibinfo {author} {\bibfnamefont {O.}~\bibnamefont {Grisel}}, \bibinfo {author} {\bibfnamefont {M.}~\bibnamefont {Blondel}}, \bibinfo {author} {\bibfnamefont {P.}~\bibnamefont {Prettenhofer}}, \bibinfo {author} {\bibfnamefont {R.}~\bibnamefont {Weiss}}, \bibinfo {author} {\bibfnamefont {V.}~\bibnamefont {Dubourg}}, \bibinfo {author} {\bibfnamefont {J.}~\bibnamefont {Vanderplas}}, \bibinfo {author} {\bibfnamefont {A.}~\bibnamefont {Passos}}, \bibinfo {author} {\bibfnamefont {D.}~\bibnamefont {Cournapeau}}, \bibinfo {author} {\bibfnamefont {M.}~\bibnamefont {Brucher}}, \bibinfo {author} {\bibfnamefont {M.}~\bibnamefont {Perrot}},\ and\ \bibinfo {author} {\bibfnamefont
  {E.}~\bibnamefont {Duchesnay}},\ }\bibfield  {title} {\bibinfo {title} {Scikit-learn: Machine learning in {P}ython},\ }\href@noop {} {\bibfield  {journal} {\bibinfo  {journal} {Journal of Machine Learning Research}\ }\textbf {\bibinfo {volume} {12}},\ \bibinfo {pages} {2825} (\bibinfo {year} {2011})}\BibitemShut {NoStop}%
\bibitem [{\citenamefont {Kingma}\ and\ \citenamefont {Ba}(2017)}]{kingma2017adammethodstochasticoptimization}%
  \BibitemOpen
  \bibfield  {author} {\bibinfo {author} {\bibfnamefont {D.~P.}\ \bibnamefont {Kingma}}\ and\ \bibinfo {author} {\bibfnamefont {J.}~\bibnamefont {Ba}},\ }\href {https://arxiv.org/abs/1412.6980} {\bibinfo {title} {Adam: A method for stochastic optimization}} (\bibinfo {year} {2017}),\ \Eprint {https://arxiv.org/abs/1412.6980} {arXiv:1412.6980 [cs.LG]} \BibitemShut {NoStop}%
\end{thebibliography}%

\end{document}